# Even faster integer multiplication

David Harvey

School of Mathematics and Statistics
University of New South Wales
Sydney NSW 2052
Australia

*Email:* d.harvey@unsw.edu.au

Joris van der Hoeven[a], Grégoire Lecerf[b]

CNRS, Laboratoire d'informatique
École polytechnique
91128 Palaiseau Cedex
France

a. *Email:* vdhoeven@lix.polytechnique.fr
b. *Email:* lecerf@lix.polytechnique.fr

July 11, 2014

We give a new proof of Fürer's bound for the cost of multiplying $n$-bit integers in the bit complexity model. Unlike Fürer, our method does not require constructing special coefficient rings with "fast" roots of unity. Moreover, we prove the more explicit bound $O(n \log n \, K^{\log^* n})$ with $K = 8$. We show that an optimised variant of Fürer's algorithm achieves only $K = 16$, suggesting that the new algorithm is faster than Fürer's by a factor of $2^{\log^* n}$. Assuming standard conjectures about the distribution of Mersenne primes, we give yet another algorithm that achieves $K = 4$.

## 1. Introduction

Let $\mathsf{I}(n)$ denote the cost of multiplying two $n$-bit integers in the deterministic multitape Turing model [38] (commonly called "bit complexity"). Previously, the best known asymptotic bound for $\mathsf{I}(n)$ was due to Fürer [18, 19]. He proved that there is a constant $K > 1$ such that

$$\mathsf{I}(n) = O(n \log n \, K^{\log^* n}), \tag{1.1}$$

where $\log^* x$, for $x \in \mathbb{R}$, denotes the iterated logarithm, i.e.,

$$\begin{aligned} \log^* x &:= \min\{k \in \mathbb{N} : \log^{\circ k} x \leqslant 1\}, \\ \log^{\circ k} &:= \underbrace{\log \circ \cdots \circ \log}_{k \times}. \end{aligned} \tag{1.2}$$

The main contribution of this paper is a new algorithm that yields the following improvement.

THEOREM 1.1. *For $n \to \infty$ we have*

$$\mathsf{I}(n) = O(n \log n \, 8^{\log^* n}).$$

Fürer suggested several methods to minimise the value of $K$ in his algorithm, but did not give an explicit bound for $K$. In section 7 of this paper, we outline an optimised variant of Fürer's algorithm that achieves $K = 16$. We do not know how to obtain $K < 16$ using Fürer's approach. This suggests that the new algorithm is faster than Fürer's by a factor of $2^{\log^* n}$.

The idea of the new algorithm is remarkably simple. Given two $n$-bit integers, we split them into chunks of exponentially smaller size, say around $\log n$ bits, and thus reduce to the problem of multiplying integer polynomials of degree $O(n/\log n)$ with coefficients of bit size $O(\log n)$. We multiply the polynomials using discrete Fourier transforms (DFTs) over $\mathbb{C}$, with a working precision of $O(\log n)$ bits. To compute the DFTs, we decompose them into "short transforms" of exponentially smaller length, say length around $\log n$, using the Cooley–Tukey method. We then use Bluestein's chirp transform to convert each short transform into a polynomial multiplication problem over $\mathbb{C}$, and finally convert back to integer multiplication via Kronecker substitution. These much smaller integer multiplications are handled recursively.

The algorithm just sketched leads immediately to a bound of the form (1.1). A detailed proof is given in section 4. We emphasise that the new method works directly over $\mathbb{C}$, and does not need special coefficient rings with "fast" roots of unity, of the type constructed by Fürer. Optimising parameters and keeping careful track of constants leads to Theorem 1.1, which is proved in section 6. We also prove the following conditional result in section 9.

THEOREM 1.2. *Assume Conjecture 9.1. Then*

$$\mathsf{I}(n) = O(n \log n \, 4^{\log^* n}).$$

Conjecture 9.1 is a slight weakening of the Lenstra–Pomerance–Wagstaff conjecture on the distribution of Mersenne primes, i.e., primes of the form $p = 2^q - 1$. The idea of the algorithm is to replace the coefficient ring $\mathbb{C}$ by the finite field $\mathbb{F}_p[\mathrm{i}]$; we are then able to exploit fast algorithms for multiplication modulo numbers of the form $2^q - 1$.

An important feature of the new algorithms is that the same techniques are applicable in other contexts, such as polynomial multiplication over finite fields. Previously, no Fürer-type complexity bounds were known for the latter problem. The details are presented in the companion paper [24].

In the remainder of this section, we present a brief history of complexity bounds for integer multiplication, and we give an overview of the paper and of our contribution. More historical details can be found in books such as [21, Chapter 8].

## 1.1. Brief history and related work

Multiplication algorithms of complexity $O(n^2)$ in the number of digits $n$ were already known in ancient civilisations. The Egyptians used an algorithm based on repeated doublings and additions. The Babylonians invented the positional numbering system, while performing their computations in base 60 instead of 10. Precise descriptions of multiplication methods close to the ones that we learn at school appeared in Europe during the late Middle Ages. For historical references, we refer to [49, Section II.5] and [37, 5].

The first subquadratic algorithm for integer multiplication, with complexity $O(n^{\log 3 / \log 2})$, was discovered by Karatsuba [30, 31]. From a modern viewpoint, Karatsuba's algorithm utilises an evaluation-interpolation scheme. The input integers are cut into smaller chunks, which are taken to be the coefficients of two integer polynomials; the polynomials are evaluated at several well-chosen points; their values at those points are (recursively) multiplied; interpolating the results at those points yields the product polynomial; finally, the integer product is recovered by pasting together the coefficients of the product polynomial. This cutting-and-pasting procedure is sometimes known as Kronecker segmentation (see section 2.6).

Shortly after the discovery of Karatsuba's algorithm, which uses three evaluation points, Toom generalised it so as to use $2r - 1$ evaluation points instead [51, 50], for any $r \geqslant 2$. This leads to the bound $\mathsf{I}(n) = O(n^{\log(2r-1)/\log r})$ for fixed $r$. Letting $r$ grow slowly with $n$, he also showed that $\mathsf{I}(n) = O(n \, 2^{5\sqrt{\log n / \log 2}})$. The algorithm was adapted to the Turing model by Cook [10] and is now known as Toom–Cook multiplication. Schönhage obtained a slightly better bound [45] by working modulo several numbers of the form $2^k - 1$ instead of using several polynomial evaluation points. Knuth proved that an even better complexity bound could be achieved by suitably adapting Toom's method [32].

The next step towards even faster integer multiplication was the rediscovery of the fast Fourier transform (FFT) by Cooley and Tukey [11] (essentially the same algorithm was already known to Gauss [27]). The FFT yields particularly efficient algorithms for evaluating and interpolating polynomials on certain special sets of evaluation points. For example, if $R$ is a ring in which 2 is invertible, and if $\omega \in R$ is a principal $2^k$-th root of unity (see section 2.2 for detailed definitions), then the FFT permits evaluation and interpolation at the points $1, \omega, ..., \omega^{2^k-1}$ using only $O(k \, 2^k)$ ring operations in $R$. Consequently, if $P$ and $Q$ are polynomials in $R[X]$ whose product has degree less than $2^k$, then the product $PQ$ can be computed using $O(k \, 2^k)$ ring operations as well.

In [47], Schönhage and Strassen presented two FFT-based algorithms for integer multiplication. In both algorithms, they first use Kronecker segmentation to convert the problem to multiplication of integer polynomials. They then embed these polynomials into $R[X]$ for a suitable ring $R$ and multiply the polynomials by using FFTs over $R$. The first algorithm takes $R = \mathbb{C}$ and $\omega = \exp(2\pi i/2^k)$, and works with finite-precision approximations to elements of $\mathbb{C}$. Multiplications in $\mathbb{C}$ itself are handled recursively, by treating them as integer multiplications (after appropriate scaling). The second algorithm, popularly known as *the* Schönhage–Strassen algorithm, takes $R = \mathbb{Z}/m\mathbb{Z}$ where $m = 2^{2^k} + 1$ is a Fermat number. This algorithm is the faster of the two, achieving the bound $\mathsf{I}(n) = O(n \log n \log \log n)$. It benefits from the fact that $\omega = 2$ is a principal $2^{k+1}$-th root of unity in $R$, and that multiplications by powers of $\omega$ can be carried out efficiently, as they correspond to simple shifts and negations. At around the same time, Pollard pointed out that one can also work with $R = \mathbb{Z}/m\mathbb{Z}$ where $m$ is a prime of the form $m = a\,2^k + 1$, since then $R^*$ contains primitive $2^k$-th roots of unity [39] (although he did not give a bound for $\mathsf{I}(n)$).

Schönhage and Strassen's algorithm remained the champion for more than thirty years, but was recently superseded by Fürer's algorithm [18]. In short, Fürer managed to combine the advantages of the two algorithms from [47], to achieve the bound $\mathsf{I}(n) = O(n \log n\, 2^{O(\log^* n)})$. Fürer's algorithm is based on the ingenious observation that the ring $R = \mathbb{C}[X]/(X^{2^{r-1}} + 1)$ contains a small number of "fast" principal $2^r$-th roots of unity, namely the powers of $X$, but also a large supply of much higher-order roots of unity inherited from $\mathbb{C}$. To evaluate an FFT over $R$, he decomposes it into many "short" transforms of length at most $2^r$, using the Cooley–Tukey method. He evaluates the short transforms with the fast roots of unity, pausing occasionally to perform "slow" multiplications by higher-order roots of unity ("twiddle factors"). A slightly subtle point of the construction is that we really need, for large $k$, a principal $2^k$-th root of unity $\omega \in R$ such that $\omega^{2^{k-r}} = X$.

In [15] it was shown that the technique from [39] to compute modulo suitable prime numbers of the form $m = a\,2^k + 1$ can be adapted to Fürer's algorithm. Although the complexity of this algorithm is essentially the same as that of Fürer's algorithm, this method has the advantage that it does not require any error analysis for approximate numerical operations in $\mathbb{C}$.

| Date | Authors | Time complexity |
| --- | --- | --- |
| <3000 BC | Unknown [37] | $O(n^2)$ |
| 1962 | Karatsuba [30, 31] | $O(n^{\log 3/\log 2})$ |
| 1963 | Toom [51, 50] | $O(n\, 2^{5\sqrt{\log n/\log 2}})$ |
| 1966 | Schönhage [45] | $O(n\, 2^{\sqrt{2\log n/\log 2}} (\log n)^{3/2})$ |
| 1969 | Knuth [32] | $O(n\, 2^{\sqrt{2\log n/\log 2}} \log n)$ |
| 1971 | Schönhage–Strassen [47] | $O(n \log n \log \log n)$ |
| 2007 | Fürer [18] | $O(n \log n\, 2^{O(\log^* n)})$ |
| 2014 | This paper | $O(n \log n\, 8^{\log^* n})$ |

**Table 1.1.** Historical overview of known complexity bounds for $n$-bit integer multiplication.

## 1.2. Our contributions and outline of the paper

Throughout the paper, integers are assumed to be handled in the standard binary representation. For our computational complexity results, we assume that we work on a Turing machine with a finite but sufficiently large number of tapes [38]. The Turing machine model is very conservative with respect to the cost of memory access, which is pertinent from a practical point of view for implementations of FFT algorithms. Nevertheless, other models for sequential computations could be considered [46, 20]. For practical purposes, parallel models might be more appropriate, but we will not consider these in this paper. Occasionally, for polynomial arithmetic over abstract rings, we will also consider algebraic complexity measures [8, Chapter 4].

In section 2, we start by recalling several classical techniques for completeness and later use: sorting and array transposition algorithms, discrete Fourier transforms (DFTs), the Cooley–Tukey algorithm, FFT multiplication and convolution, Bluestein's chirp transform, and Kronecker substitution and segmentation. In section 3, we also provide the necessary tools for the error analysis of complex Fourier transforms. Most of these tools are standard, although our presentation is somewhat *ad hoc*, being based on fixed point arithmetic.

In section 4, we describe a simplified version of the new integer multiplication algorithm, without any attempt to minimise the aforementioned constant $K$. As mentioned in the sketch above, the key idea is to reduce a given DFT over $\mathbb{C}$ to a collection of "short" transforms, and then to convert these short transforms back to integer multiplication by a combination of Bluestein's chirp transform and Kronecker substitution.

The complexity analysis of Fürer's algorithm and the algorithm from section 4 involves functional inequalities which contain post-compositions with logarithms and other slowly growing functions. In section 5, we present a few systematic tools for analysing these types of inequalities. For more information on this quite particular kind of asymptotic analysis, we refer the reader to [44, 16].

In section 6, we present an optimised version of the algorithm from section 4, proving in particular the bound $\mathsf{I}(n) = O(n \log n\, 8^{\log^* n})$ (Theorem 1.1), which constitutes the main result of this paper. In section 7, we outline a similar complexity analysis for Fürer's algorithm. Even after several optimisations of the original algorithm, we were unable to attain a bound better than $\mathsf{I}(n) = O(n \log n\, 16^{\log^* n})$. This suggests that the new algorithm outperforms Fürer's algorithm by a factor of $2^{\log^* n}$.

This speedup is surprising, given that the short transforms in Fürer's algorithm involve only shifts, additions and subtractions. The solution to the paradox is that Fürer has made the short transforms *too fast*. Indeed, they are so fast that they make a negligible contribution to the overall complexity, and his computation is dominated by the "slow" twiddle factor multiplications. In the new algorithm, we push more work into the short transforms, allowing them to get slightly slower; the *quid pro quo* is that we avoid the factor of two in zero-padding caused by Fürer's introduction of artificial "fast" roots of unity. The optimal strategy is actually to let the short transforms dominate the computation, by increasing the short transform length relative to the coefficient size. Fürer is unable to do this, because in his algorithm these two parameters are too closely linked. To underscore just how far the situation has been inverted relative to Fürer's algorithm, we point out that in our presentation we can get away with using Schönhage–Strassen for the twiddle factor multiplications, without any detrimental effect on the overall complexity.

We have chosen to base most of our algorithms on approximate complex arithmetic. Instead, following [39] and [15], we might have chosen to use modular arithmetic. In section 8, we will briefly indicate how our main algorithm can be adapted to this setting. This variant of our algorithm presents several analogies with its adaptation to polynomial multiplication over finite fields [24].

The question remains whether there exists an even faster algorithm than the algorithm of section 6. In an earlier paper [17], Fürer gave another algorithm of complexity $O(n \log n\, 2^{O(\log^* n)})$ under the assumption that there exist sufficiently many Fermat primes, i.e., primes of the form $F_m = 2^{2^m} + 1$. It can be shown that a careful optimisation of this algorithm yields the bound $\mathsf{I}(n) = O(n \log n\, 4^{\log^* n})$. Unfortunately, odds are high that $F_4$ is the largest Fermat prime. In section 9, we present an algorithm that achieves the bound $\mathsf{I}(n) = O(n \log n\, 4^{\log^* n})$ under the more plausible conjecture that there exist sufficiently many Mersenne primes (Theorem 1.2). The main technical ingredient is a variant of an algorithm of Crandall and Fagin [12] that permits efficient multiplication modulo $2^q - 1$, despite $q$ not being divisible by a large power of two.

It would be interesting to know whether the new algorithms could be useful in practice. We have implemented an unoptimised version of the algorithm from section 8 in the MATHEMAGIX system [29] and found our implementation to be an order of magnitude slower than the GMP library [23]. There is certainly room for improvement, but we doubt that even a highly optimised implementation of the new algorithm will be competitive in the near future. Nevertheless, the variant for polynomial multiplication over finite fields presented in [24] seems to be a promising avenue for achieving speedups in practical computations. This will be investigated in a forthcoming paper.

**Notations.** We use Hardy's notations $f \prec g$ for $f = o(g)$, and $f \asymp g$ for $f = O(g)$ and $g = O(f)$. The symbol $\mathbb{R}^{\geqslant}$ denotes the set of non-negative real numbers, and $\mathbb{N}$ denotes $\{0, 1, 2, ...\}$. We will write $\lg n := \lceil \log n / \log 2 \rceil$.

## 2. Survey of classical tools

This section recalls basic facts on Fourier transforms and related techniques used in subsequent sections. For more details and historical references we refer the reader to standard books on the subject such as [2, 8, 21, 42].

### 2.1. Arrays and sorting

In the Turing model, we have available a fixed number of linear tapes. An $n_1 \times \cdots \times n_d$ array $M_{i_1,...,i_d}$ of $b$-bit elements is stored as a linear array of $n_1 \cdots n_d b$ bits. We generally assume that the elements are ordered lexicographically by $(i_1, ..., i_d)$, though this is just an implementation detail.

What is significant from a complexity point of view is that occasionally we must switch representations, to access an array (say 2-dimensional) by "rows" or by "columns". In the Turing model, we may transpose an $n_1 \times n_2$ matrix of $b$-bit elements in time $O(b\, n_1\, n_2 \lg \min(n_1, n_2))$, using the algorithm of [4, Appendix]. Briefly, the idea is to split the matrix into two halves along the "short" dimension, and transpose each half recursively.

We will also require more complex rearrangements of data, for which we resort to sorting. Suppose that $X$ is a totally ordered set, whose elements are represented by bit strings of length $b$, and suppose that we can compare elements of $X$ in time $O(b)$. Then an array of $n$ elements of $X$ may be sorted in time $O(b\, n \lg n)$ using merge sort [33], which can be implemented efficiently on a Turing machine.

### 2.2. Discrete Fourier transforms

Let $R$ be a commutative ring with identity and let $n \geqslant 1$. An element $\omega \in R$ is said to be a *principal $n$-th root of unity* if $\omega^n = 1$ and

$$\sum_{k=0}^{n-1} (\omega^i)^k = 0 \tag{2.1}$$

for all $i \in \{1, ..., n-1\}$. In this case, we define the *discrete Fourier transform* (or DFT) of an $n$-tuple $a = (a_0, ..., a_{n-1}) \in R^n$ with respect to $\omega$ to be $\mathrm{DFT}_\omega(a) = \hat{a} = (\hat{a}_0, ..., \hat{a}_{n-1}) \in R^n$ where

$$\hat{a}_i := a_0 + a_1 \omega^i + \cdots + a_{n-1} \omega^{(n-1)i}.$$

That is, $\hat{a}_i$ is the evaluation of the polynomial $A(X) := a_0 + a_1 X + \cdots + a_{n-1} X^{n-1}$ at $\omega^i$.

If $\omega$ is a principal $n$-th root of unity, then so is its inverse $\omega^{-1} = \omega^{n-1}$, and we have

$$\mathrm{DFT}_{\omega^{-1}}(\mathrm{DFT}_\omega(a)) = n\, a.$$

Indeed, writing $b := \mathrm{DFT}_{\omega^{-1}}(\mathrm{DFT}_\omega(a))$, the relation (2.1) implies that

$$b_i = \sum_{j=0}^{n-1} \hat{a}_j \omega^{-ji} = \sum_{j=0}^{n-1} \sum_{k=0}^{n-1} a_k \omega^{j(k-i)} = \sum_{k=0}^{n-1} a_k \sum_{j=0}^{n-1} \omega^{j(k-i)} = \sum_{k=0}^{n-1} a_k (n\, \delta_{i,k}) = n\, a_i,$$

where $\delta_{i,k} = 1$ if $i = k$ and $\delta_{i,k} = 0$ otherwise.

**Remark 2.1.** In all of the new algorithms introduced in this paper, we actually work over a field, whose characteristic does not divide $n$. In this setting, the concept of principal root of unity coincides with the more familiar *primitive root of unity*. The more general "principal root" concept is only needed for discussions of other algorithms, such as the Schönhage–Strassen algorithm or Fürer's algorithm.

## 2.3. The Cooley–Tukey FFT

Let $\omega$ be a principal $n$-th root of unity and let $n = n_1 n_2$ where $1 < n_1 < n$. Then $\omega^{n_1}$ is a principal $n_2$-th root of unity and $\omega^{n_2}$ is a principal $n_1$-th root of unity. Moreover, for any $i_1 \in \{0, ..., n_1 - 1\}$ and $i_2 \in \{0, ..., n_2 - 1\}$, we have

$$\begin{aligned} \hat{a}_{i_1 n_2 + i_2} &= \sum_{k_1=0}^{n_1-1} \sum_{k_2=0}^{n_2-1} a_{k_2 n_1 + k_1} \omega^{(k_2 n_1 + k_1)(i_1 n_2 + i_2)} \\ &= \sum_{k_1=0}^{n_1-1} \omega^{k_1 i_2} \left( \sum_{k_2=0}^{n_2-1} a_{k_2 n_1 + k_1} (\omega^{n_1})^{k_2 i_2} \right) (\omega^{n_2})^{k_1 i_1}. \end{aligned} \quad (2.2)$$

If $\mathcal{A}_1$ and $\mathcal{A}_2$ are algorithms for computing DFTs of length $n_1$ and $n_2$, we may use (2.2) to construct an algorithm $\mathcal{A}_1 \odot \mathcal{A}_2$ for computing DFTs of length $n$ as follows.

For each $k_1 \in \{0, ..., n_1 - 1\}$, the sum inside the brackets corresponds to the $i_2$-th coefficient of a DFT of the $n_2$-tuple $(a_{0 n_1 + k_1}, ..., a_{(n_2 - 1) n_1 + k_1}) \in R^{n_2}$ with respect to $\omega^{n_1}$. Evaluating these *inner DFTs* requires $n_1$ calls to $\mathcal{A}_2$. Next, we multiply by the *twiddle factors* $\omega^{k_1 i_2}$, at a cost of $n$ operations in $R$. (Actually, fewer than $n$ multiplications are required, as some of the twiddle factors are equal to 1. This optimisation, while important in practice, has no asymptotic effect on the algorithms discussed in this paper.) Finally, for each $i_2 \in \{0, ..., n_2 - 1\}$, the outer sum corresponds to the $i_1$-th coefficient of a DFT of an $n_1$-tuple in $R^{n_1}$ with respect to $\omega^{n_2}$. These *outer DFTs* require $n_2$ calls to $\mathcal{A}_1$.

Denoting by $\mathsf{F}_R(n)$ the number of ring operations needed to compute a DFT of length $n$, and assuming that we have available a precomputed table of twiddle factors, we obtain

$$\mathsf{F}_R(n_1 n_2) \leqslant n_1 \mathsf{F}_R(n_2) + n_2 \mathsf{F}_R(n_1) + n.$$

For a factorisation $n = n_1 \cdots n_d$, this yields recursively

$$\mathsf{F}_R(n) \leqslant \sum_{i=1}^{d} \frac{n}{n_i} \mathsf{F}_R(n_i) + (d-1) n. \quad (2.3)$$

The corresponding algorithm is denoted $\mathcal{A}_1 \odot \cdots \odot \mathcal{A}_d$. The $\odot$ operation is neither commutative nor associative; the above expression will always be taken to mean $(\cdots((\mathcal{A}_1 \odot \mathcal{A}_2) \odot \mathcal{A}_3) \odot \cdots) \odot \mathcal{A}_d$.

Let $\mathcal{B}$ be the butterfly algorithm that computes a DFT of length 2 by the formula $(a_0, a_1) \mapsto (a_0 + a_1, a_0 - a_1)$. Then $\mathcal{B}^{\odot k} := \mathcal{B} \odot \cdots \odot \mathcal{B}$ computes a DFT of length $n := 2^k$ in time $\mathsf{F}_R(2^k) = O(k n)$. Algorithms of this type are called *fast Fourier transforms* (or FFTs).

The above discussion requires several modifications in the Turing model. Assume that elements of $R$ are represented by $b$ bits.

First, for $\mathcal{A}_1 \odot \mathcal{A}_2$, we must add a rearrangement cost of $O(b n \lg \min(n_1, n_2))$ to efficiently access the rows and columns for the recursive subtransforms (see section 2.1). For the general case $\mathcal{A}_1 \odot \cdots \odot \mathcal{A}_d$, the total rearrangement cost is bounded by $O(\sum_i b n \lg n_i) = O(b n \lg n)$.

Second, we will sometimes use *non-algebraic* algorithms to compute the subtransforms, so it may not make sense to express their cost in terms of $\mathsf{F}_R$. The relation (2.3) therefore becomes

$$\mathsf{F}(n) \leqslant \sum_{i=1}^{d} \frac{n}{n_i} \mathsf{F}(n_i) + (d-1) n \, \mathsf{m}_R + O(b n \lg n), \quad (2.4)$$

where $\mathsf{F}(n)$ is the (Turing) cost of a transform of length $n$ over $R$, and where $\mathsf{m}_R$ is the cost of a single multiplication in $R$.

Finally, we point out that $\mathcal{A}_1 \odot \mathcal{A}_2$ requires access to a table of twiddle factors $\omega^{i_1 i_2}$, ordered lexicographically by $(i_1, i_2)$, for $0 \leqslant i_1 < n_1$, $0 \leqslant i_2 < n_2$. Assuming that we are given as input a precomputed table of the form $1, \omega, ..., \omega^{n-1}$, we must show how to extract the required twiddle factor table in the correct order. We first construct a list of triples $(i_1, i_2, i_1 i_2)$, ordered by $(i_1, i_2)$, in time $O(n \lg n)$; then sort by $i_1 i_2$ in time $O(n \lg^2 n)$ (see section 2.1); then merge with the given root table to obtain a table $(i_1, i_2, \omega^{i_1 i_2})$, ordered by $i_1 i_2$, in time $O(n (b + \lg n))$; and finally sort by $(i_1, i_2)$ in time $O(n \lg n (b + \lg n))$. The total cost of the extraction is thus $O(n \lg n (b + \lg n))$.

The corresponding cost for $\mathcal{A}_1 \odot \cdots \odot \mathcal{A}_d$ is determined as follows. Assuming that the table $1, \omega, \ldots, \omega^{n-1}$ is given as input, we first extract the subtables of $(n_1 \cdots n_i)$-th roots of unity for $i = d-1, \ldots, 2$ in time $O((n_1 \cdots n_d + \cdots + n_1 n_2)(b + \lg n)) = O(n(b + \lg n))$. Extracting the twiddle factor table for the decomposition $(n_1 \cdots n_{i-1}) \times n_i$ then costs $O(n_1 \cdots n_i \lg n (b + \lg n))$; the total over all $i$ is again $O(n \lg n (b + \lg n))$.

**Remark 2.2.** An alternative approach is to compute the twiddle factors directly in the correct order. When working over $\mathbb{C}$, as in section 3, this requires a slight increase in the working precision. Similar comments apply to the root tables used in Bluestein's algorithm in section 2.5.

## 2.4. Fast Fourier multiplication

Let $\omega$ be a principal $n$-th root of unity in $R$ and assume that $n$ is invertible in $R$. Consider two polynomials $A = a_0 + \cdots + a_{n-1} X^{n-1}$ and $B = b_0 + \cdots + b_{n-1} X^{n-1}$ in $R[X]$. Let $C = c_0 + \cdots + c_{n-1} X^{n-1}$ be the polynomial defined by

$$c := \tfrac{1}{n} \mathrm{DFT}_{\omega^{-1}}(\mathrm{DFT}_\omega(a) \, \mathrm{DFT}_\omega(b)),$$

where the product of the DFTs is taken pointwise. By construction, we have $\hat{c} = \hat{a} \, \hat{b}$, which means that $C(\omega^i) = A(\omega^i) B(\omega^i)$ for all $i \in \{0, \ldots, n-1\}$. The product $S = s_0 + \cdots + s_{n-1} X^{n-1}$ of $A$ and $B$ modulo $X^n - 1$ also satisfies $S(\omega^i) = A(\omega^i) B(\omega^i)$ for all $i$. Consequently, $\hat{s} = \hat{a} \, \hat{b}$, $s = \mathrm{DFT}_{\omega^{-1}}(\hat{s})/n = c$, whence $C = S$.

For polynomials $A, B \in R[X]$ with $\deg A < n$ and $\deg B < n$, we thus obtain an algorithm for the computation of $A B$ modulo $X^n - 1$ using at most $3 \, \mathsf{F}_R(n) + O(n)$ operations in $R$. Modular products of this type are also called *cyclic convolutions*. If $\deg (A B) < n$, then we may recover the product $A B$ from its reduction modulo $X^n - 1$. This multiplication method is called *FFT multiplication*.

If one of the arguments (say $B$) is fixed and we want to compute many products $A B$ (or cyclic convolutions) for different $A$, then we may precompute $\mathrm{DFT}_\omega(b)$, after which each new product $A B$ can be computed using only $2 \, \mathsf{F}_R(n) + O(n)$ operations in $R$.

## 2.5. Bluestein's chirp transform

We have shown above how to multiply polynomials using DFTs. Inversely, it is possible to reduce the computation of DFTs — of arbitrary length, not necessarily a power of two — to polynomial multiplication [3], as follows.

Let $\omega$ be a principal $n$-th root of unity. For simplicity we assume that $n$ is even, and that there exists some $\eta \in R$ with $\eta^2 = \omega$. Consider the sequences

$$f_i := \eta^{i^2}, \quad g_i := \eta^{-i^2}.$$

Then $\omega^{ij} = f_i f_j g_{i-j}$, so for any $a \in R^n$ we have

$$\hat{a}_i = \sum_{j=0}^{n-1} a_j \omega^{ij} = f_i \sum_{j=0}^{n-1} (a_j f_j) g_{i-j}. \tag{2.5}$$

Also, since $n$ is even,

$$g_{i+n} = \eta^{-(i+n)^2} = \eta^{-i^2 - n^2 - 2ni} = \eta^{-i^2} \omega^{-(\frac{n}{2}+i)n} = g_i.$$

Now let $F := f_0 a_0 + \cdots + f_{n-1} a_{n-1} X^{n-1}$, $G := g_0 + \cdots + g_{n-1} X^{n-1}$ and $C := c_0 + \cdots + c_{n-1} X^{n-1} \equiv F G$ modulo $X^n - 1$. Then (2.5) implies that $\hat{a}_i = f_i c_i$ for all $i \in \{0, \ldots, n-1\}$. In other words, the computation of a DFT of even length $n$ reduces to a cyclic convolution product of the same length, together with $O(n)$ additional operations in $R$. Notice that the polynomial $G$ is fixed and independent of $a$ in this product.

The only complication in the Turing model is the cost of extracting the $f_i$ in the correct order, i.e., in the order $1, \eta, \eta^4, \eta^9, \ldots, \eta^{(n-1)^2}$, given as input a precomputed table $1, \eta, \eta^2, \ldots, \eta^{2n-1}$. We may do this in time $O(n \lg n (b + \lg n))$ by applying the strategy from section 2.3 to the pairs $(i, i^2 \bmod 2n)$ for $0 \leqslant i < n$. Similar remarks apply to the $g_i$.

**Remark 2.3.** It is also possible to give variants of the new multiplication algorithms in which Bluestein's transform is replaced by a different method for converting DFTs to convolutions, such as Rader's algorithm [41].

## 2.6. Kronecker substitution and segmentation

Multiplication in $\mathbb{Z}[X]$ may be reduced to multiplication in $\mathbb{Z}$ using the classical technique of *Kronecker substitution* [21, Corollary 8.27]. More precisely, let $d > 0$ and $n > 0$, and suppose that we are given two polynomials $A, B \in \mathbb{Z}[X]$ of degree less than $d$, with coefficients $A_i$ and $B_i$ satisfying $|A_i| \leqslant 2^n$ and $|B_i| \leqslant 2^n$. Then for the product $C = AB$ we have $|C_i| \leqslant 2^{2n + \lg d}$. Consequently, the coefficients of $C$ may be read off the integer product $C(2^N) = A(2^N) B(2^N)$ where $N := 2n + \lg d + 2$. Notice that the integers $|A(2^N)|$ and $|B(2^N)|$ have bit length at most $dN$, and the encoding and decoding processes have complexity $O(dN)$.

The inverse procedure is *Kronecker segmentation*. Given $n > 0$ and $d > 0$, and non-negative integers $a < 2^n$ and $b < 2^n$, we may reduce the computation of $c := ab$ to the computation of a product $C := AB$ of two polynomials $A, B \in \mathbb{Z}[X]$ of degree less than $d$, and with $|A_i| < 2^k$ and $|B_i| < 2^k$ where $k := \lceil n/d \rceil$. Indeed, we may cut the integers into chunks of $k$ bits each, so that $a = A(2^k)$, $b = B(2^k)$ and $c = C(2^k)$. Notice that we may recover $c$ from $C$ using an overlap-add procedure in time $O(d(k + \lg d)) = O(n + d \lg d)$. In our applications, we will always have $d = O(n/\lg n)$, so that $O(n + d \lg d) = O(n)$.

Kronecker substitution and segmentation can also be used to handle Gaussian integers (and Gaussian integer polynomials), and to compute cyclic convolutions. For example, given polynomials $A, B \in \mathbb{Z}[\mathrm{i}][X]/(X^d - 1)$ with $|A_i|, |B_i| \leqslant 2^n$, then for $C = AB$ we have $|C_i| \leqslant 2^{2n + \lg d}$, so we may recover $C$ from the cyclic Gaussian integer product $C(2^N) = A(2^N) B(2^N) \in (\mathbb{Z}/(2^{dN} - 1)\mathbb{Z})[\mathrm{i}]$, where $N := 2n + \lg d + 2$. In the other direction, suppose that we wish to compute $ab$ for some $a, b \in (\mathbb{Z}/(2^{dn} - 1)\mathbb{Z})[\mathrm{i}]$. We may assume that the "real" and "imaginary" parts of $a$ and $b$ are non-negative, and so reduce to the problem of multiplying $A, B \in \mathbb{Z}[\mathrm{i}][X]/(X^d - 1)$, where $a = A(2^n)$ and $b = B(2^n)$, and where the real and imaginary parts of $A_i, B_i \in \mathbb{Z}[\mathrm{i}]$ are non-negative and have at most $n$ bits.

## 3. FIXED POINT COMPUTATIONS AND ERROR BOUNDS

In this section, we consider the computation of DFTs over $\mathbb{C}$ in the Turing model. Elements of $\mathbb{C}$ can only be represented approximately on a Turing machine. We describe algorithms that compute DFTs approximately, using a fixed-point representation for $\mathbb{C}$, and we give complexity bounds and a detailed error analysis for these algorithms. We refer the reader to [7] for more details about multiple precision arithmetic.

For our complexity estimates we will freely use the standard observation that $\mathsf{I}(O(n)) = O(\mathsf{I}(n))$, since the multiplication of two integers of bit length $\leqslant kn$ reduces to $k^2$ multiplications of integers of bit length $\leqslant n$, for any fixed $k \geqslant 1$.

### 3.1. Fixed point numbers

We will represent fixed point numbers by a signed mantissa and a fixed exponent. More precisely, given a precision parameter $p \geqslant 4$, we denote by $\mathbb{C}_p$ the set of complex numbers of the form $z = m_z 2^{-p}$, where $m_z = u + v\mathrm{i}$ for integers $u$ and $v$ satisfying $u^2 + v^2 \leqslant 2^{2p}$, i.e., $|z| \leqslant 1$. We write $\mathbb{C}_p 2^e$ for the set of complex numbers of the form $u 2^e$, where $u \in \mathbb{C}_p$ and $e \in \mathbb{Z}$; in particular, for $z \in \mathbb{C}_p 2^e$ we always have $|z| \leqslant 2^e$. At every stage of our algorithms, the exponent $e$ will be determined implicitly by context, and in particular, the exponents do not have to be explicitly stored or manipulated.

In our error analysis of numerical algorithms, each $z \in \mathbb{C}_p 2^e$ is really the approximation of some genuine complex number $\tilde{z} \in \mathbb{C}$. Each such $z$ comes with an implicit error bound $\varepsilon_z \geqslant 0$; this is a real number for which we can guarantee that $|z - \tilde{z}| \leqslant \varepsilon_z$. We also define the relative error bound for $z$ by $\rho_z := \varepsilon_z / 2^e$. We finally denote by $\epsilon := 2^{1-p} \leqslant 1/8$ the "machine accuracy".

**Remark 3.1.** Interval arithmetic [36] (or ball arithmetic [28, Chapter 3]) provides a systematic method for tracking error bounds by storing the bounds along with $z$. We will use similar formulas for the computation of $\varepsilon_z$ and $\rho_z$, but we will not actually store the bounds during computations.

## 3.2. Basic arithmetic

In this section we give error bounds and complexity estimates for fixed point addition, subtraction and multiplication, under certain simplifying assumptions. In particular, in our DFTs, we only ever need to add and subtract numbers with the same exponent. We also give error bounds for fixed point convolution of vectors; the complexity of this important operation is considered later.

For $x \in \mathbb{R}$, we define the "round towards zero" function $\lfloor x \rceil$ by $\lfloor x \rceil := \lfloor x \rfloor$ if $x \geqslant 0$ and $\lfloor x \rceil := \lceil x \rceil$ if $x \leqslant 0$. For $x, y \in \mathbb{R}$, we define $\lfloor x + y\, \mathrm{i} \rceil := \lfloor x \rceil + \lfloor y \rceil \mathrm{i}$. Notice that $|\lfloor z \rceil| \leqslant |z|$ and $|\lfloor z \rceil - z| \leqslant \sqrt{2}$ for any $z \in \mathbb{C}$.

**PROPOSITION 3.2.** *Let $z, u \in \mathbb{C}_p 2^e$. Define the fixed point sum and difference $z \dotplus u, z \dotminus u \in \mathbb{C}_p 2^{e+1}$ by $m_{z \pm u} := \lfloor (m_z \pm m_u)/2 \rceil$. Then $z \dotplus u$ and $z \dotminus u$ can be computed in time $O(p)$, and*

$$\rho_{z \dot\pm u} \;\leqslant\; \frac{\rho_z + \rho_u}{2} + \epsilon.$$

**Proof.** We have

$$\frac{|(z \dot\pm u) - (z \pm u)|}{2^{e+1}} = \left| \left\lfloor \frac{m_z \pm m_u}{2} \right\rceil - \frac{m_z \pm m_u}{2} \right| 2^{-p} \leqslant \sqrt{2} \cdot 2^{-p} \leqslant \epsilon$$

and

$$\frac{|(z \pm u) - (\tilde{z} \pm \tilde{u})|}{2^{e+1}} \leqslant \frac{\varepsilon_z + \varepsilon_u}{2^{e+1}} = \frac{\rho_z + \rho_u}{2},$$

whence $|(z \dot\pm u) - (\tilde{z} \pm \tilde{u})|/2^{e+1} \leqslant (\rho_z + \rho_u)/2 + \epsilon$. □

**PROPOSITION 3.3.** *Let $z \in \mathbb{C}_p 2^{e_z}$ and $u \in \mathbb{C}_p 2^{e_u}$. Define the fixed point product $z \dottimes u \in \mathbb{C}_p 2^{e_z + e_u}$ by $m_{z \dottimes u} := \lfloor 2^{-p} m_z m_u \rceil$. Then $z \dottimes u$ can be computed in time $O(\mathsf{I}(p))$, and*

$$1 + \rho_{z \dottimes u} \;\leqslant\; (1 + \rho_z)(1 + \rho_u)(1 + \epsilon).$$

**Proof.** We have

$$|z \dottimes u - z\, u|/2^{e_z + e_u} = |\lfloor 2^{-p} m_z m_u \rceil - 2^{-p} m_z m_u| \, 2^{-p} \leqslant \sqrt{2} \cdot 2^{-p} \leqslant \epsilon$$

and

$$\begin{aligned}
|z\, u - \tilde{z}\, \tilde{u}| &\leqslant |z|\,|u - \tilde{u}| + |z - \tilde{z}|\,(|u| + |\tilde{u} - u|) \\
&\leqslant 2^{e_z} \varepsilon_u + 2^{e_u} \varepsilon_z + \varepsilon_z \varepsilon_u \\
&= (\rho_u + \rho_z + \rho_z \rho_u)\, 2^{e_z + e_u}.
\end{aligned}$$

Consequently, $|z \dottimes u - \tilde{z}\, \tilde{u}|/2^{e_z + e_u} \leqslant \rho_z + \rho_u + \rho_z \rho_u + \epsilon \leqslant (1 + \rho_z)(1 + \rho_u)(1 + \epsilon) - 1$. □

Proposition 3.3 may be generalised to numerical cyclic convolution of vectors as follows.

**PROPOSITION 3.4.** *Let $k \geqslant 1$ and $n := 2^k$. Let $z \in (\mathbb{C}_p 2^{e_z})^n$ and $u \in (\mathbb{C}_p 2^{e_u})^n$. Define the fixed point convolution $z \,\mathsf{<dotast>}\, u \in (\mathbb{C}_p 2^{e_z + e_u + k})^n$ by*

$$m_{(z\,\mathsf{<dotast>}\,u)_i} := \left\lfloor 2^{-p-k} \sum_{i_1 + i_2 = i \,(\mathrm{mod}\, n)} m_{z_{i_1}} m_{u_{i_2}} \right\rceil, \qquad 0 \leqslant i < n.$$

*Then*

$$\max_i\, (1 + \rho_{(z\,\mathsf{<dotast>}\,u)_i}) \;\leqslant\; \max_i\, (1 + \rho_{z_i}) \max_i\, (1 + \rho_{u_i})(1 + \epsilon).$$

**Proof.** Let $*$ denote the exact convolution, and write $\rho_z := \max_j \rho_{z_j}$ and $\rho_u := \max_j \rho_{u_j}$. As in the proof of Proposition 3.3, we obtain $|(z \,\mathsf{<dotast>}\, u)_i - (z * u)_i|/2^{e_z + e_u + k} \leqslant \sqrt{2} \cdot 2^{-p} \leqslant \epsilon$ and

$$\begin{aligned}
|(z * u)_i - (\tilde{z} * \tilde{u})_i| &\leqslant \sum_{i_1 + i_2 = i \,(\mathrm{mod}\, n)} |z_{i_1} u_{i_2} - \tilde{z}_{i_1} \tilde{u}_{i_2}| \\
&\leqslant (\rho_z + \rho_u + \rho_z \rho_u)\, 2^{e_z + e_u + k}.
\end{aligned}$$

The proof is concluded in the same way as Proposition 3.3. □

### 3.3. Precomputing roots of unity

Let $\mathbb{H} := \{x + y\,\mathrm{i} \in \mathbb{C} : y \geqslant 0\}$ and $\mathbb{H}_p := \{x + y\,\mathrm{i} \in \mathbb{C}_p : y \geqslant 0\}$. Let $\sqrt{\ } : \mathbb{H} \to \mathbb{H}$ be the branch of the square root function such that $\sqrt{\mathrm{e}^{\mathrm{i}\theta}} := \mathrm{e}^{\mathrm{i}\theta/2}$ for $0 \leqslant \theta \leqslant \pi$. Using Newton's method [7, Section 3.5] and Schönhage–Strassen multiplication [47], we may construct a fixed point square root function $\dot{\sqrt{\ }} : \mathbb{H}_p \to \mathbb{H}_p$, which may be evaluated in time $O(p \log p \log \log p)$, such that $|\dot{\sqrt{z}} - \sqrt{z}| \leqslant \epsilon$ for all $z \in \mathbb{H}_p$. For example, we may first compute some $u \in \mathbb{H}$ such that $|u - \sqrt{z}| \leqslant \epsilon/4$ and $|u| \leqslant 1$, and then take $\dot{\sqrt{z}} := \lfloor 2^p u \rfloor 2^{-p}$; the desired bound follows since $\epsilon/4 + \sqrt{2} \cdot 2^{-p} \leqslant \epsilon$.

LEMMA 3.5. *Let $z \in \mathbb{H}_p$, and assume that $|\tilde{z}| = 1$ and $\rho_z \leqslant 3/8$. Then $\rho_{\dot{\sqrt{z}}} \leqslant \rho_z + \epsilon$.*

**Proof.** The mean value theorem implies that $|\sqrt{\tilde{z}} - \sqrt{z}| \leqslant \varepsilon_z \max_{w \in D} |1/(2\sqrt{w})|$ where $D := \{w \in \mathbb{H} : |w - z| \leqslant \varepsilon_z\}$. For $w \in D$ we have $|w| \geqslant |\tilde{z}| - |\tilde{z} - z| - |z - w| \geqslant 1 - 3/8 - 3/8 \geqslant 1/4$; hence $|\sqrt{\tilde{z}} - \sqrt{z}| \leqslant \varepsilon_z = \rho_z$. By construction $|\dot{\sqrt{z}} - \sqrt{z}| \leqslant \epsilon$. We conclude that $|\dot{\sqrt{z}} - \sqrt{\tilde{z}}| \leqslant \rho_z + \epsilon$. □

PROPOSITION 3.6. *Let $k \in \mathbb{N}$ and $p \geqslant k$, and let $\omega := \mathrm{e}^{2\pi\mathrm{i}/2^k}$. We may compute $1, \omega, \omega^2, ..., \omega^{2^k-1} \in \mathbb{C}_p$, with $\rho_{\omega^i} \leqslant \epsilon$ for all $i$, in time $O(2^k p \log p \log \log p)$.*

**Proof.** It suffices to compute $1, \omega, ..., \omega^{2^{k-1}-1} \in \mathbb{H}_p$. Starting from $\omega^0 = 1$ and $\omega^{2^{k-2}} = \mathrm{i}$, for each $\ell = k - 3, k - 4, ..., 0$, we compute $\omega^{i 2^\ell}$ for $i = 1, 3, ..., 2^{k-\ell-1} - 1$ using $\omega^{i 2^\ell} := \dot{\sqrt{\omega^{i 2^{\ell+1}}}}$ if $i < 2^{k-\ell-2}$ and $\omega^{i 2^\ell} := \mathrm{i}\,\omega^{i 2^\ell - 2^{k-2}}$ otherwise. Performing all computations with temporarily increased precision $p' := p + \lg p + 2$ and corresponding $\epsilon' := 2^{1-p'}$, Lemma 3.5 yields $\rho_{\omega^i} \leqslant k\,\epsilon' \leqslant \epsilon/4$. This also shows that the hypothesis $\rho_{\omega^i} \leqslant 3/8$ is always satisfied, since $\epsilon/4 \leqslant 1/32 \leqslant 3/8$. After rounding to $p$ bits, the relative error is at most $\epsilon/4 + \sqrt{2} \cdot 2^{-p} \leqslant \epsilon$. □

### 3.4. Error analysis for fast Fourier transforms

A *tight* algorithm for computing DFTs of length $n = 2^k \geqslant 2$ is a numerical algorithm that takes as input an $n$-tuple $a \in (\mathbb{C}_p\, 2^e)^n$ and computes an approximation $\hat{a} \in (\mathbb{C}_p\, 2^{e+k})^n$ to the DFT of $a$ with respect to $\omega = \mathrm{e}^{2\pi\mathrm{i}/n}$ (or $\omega = \mathrm{e}^{-2\pi\mathrm{i}/n}$ in the case of an inverse transform), such that

$$\max_i (1 + \rho_{\hat{a}_i}) \leqslant \max_i (1 + \rho_{a_i})(1 + \epsilon)^{3k-2}.$$

We assume for the moment that any such algorithm has at its disposal all necessary root tables with relative error not exceeding $\epsilon$. Propositions 3.2 and 3.3 directly imply the following:

PROPOSITION 3.7. *The butterfly algorithm $\mathcal{B}$ that computes a DFT of length 2 using the formula $(a_0, a_1) \mapsto (a_0 \dot{+} a_1, a_0 \dot{-} a_1)$ is tight.*

**Proof.** We have $\rho_{\hat{a}_i} \leqslant (\rho_{a_0} + \rho_{a_1})/2 + \epsilon \leqslant \max_i \rho_{a_i} + \epsilon \leqslant (1 + \max_i \rho_{a_i})(1 + \epsilon) - 1$. □

PROPOSITION 3.8. *Let $k_1, k_2 \geqslant 1$, and let $\mathcal{A}_1$ and $\mathcal{A}_2$ be tight algorithms for computing DFTs of lengths $2^{k_1}$ and $2^{k_2}$. Then $\mathcal{A}_1 \odot \mathcal{A}_2$ is a tight algorithm for computing DFTs of length $2^{k_1+k_2}$.*

**Proof.** The inner and outer DFTs contribute factors of $(1+\epsilon)^{3k_1-2}$ and $(1+\epsilon)^{3k_2-2}$, and by Proposition 3.3 the twiddle factor multiplications contribute a factor of $(1+\epsilon)^2$. Thus

$$\max_i (1 + \rho_{\hat{a}_i}) \leqslant \max_i (1 + \rho_{a_i})(1+\epsilon)^{(3k_1-2)+2+(3k_2-2)} \leqslant \max_i (1 + \rho_{a_i})(1+\epsilon)^{3(k_1+k_2)-2}. \quad \square$$

COROLLARY 3.9. *Let $k \geqslant 1$. Then $\mathcal{B}^{\odot k}$ is a tight algorithm for computing DFTs of length $2^k$ over $\mathbb{C}_p$, whose complexity is bounded by $O(2^k k\, \mathsf{I}(p))$.*

## 4. A SIMPLE AND FAST MULTIPLICATION ALGORITHM

In this section we give the simplest version of the new integer multiplication algorithm. The key innovation is an alternative method for computing DFTs of small length. This new method uses a combination of Bluestein's chirp transform and Kronecker substitution (see sections 2.5 and 2.6) to convert the DFT to a cyclic integer product in $(\mathbb{Z}/(2^{n'} - 1)\mathbb{Z})[\mathrm{i}]$ for suitable $n'$.

PROPOSITION 4.1. *Let $1 \leqslant r \leqslant p$. There exists a tight algorithm $\mathcal{C}_r$ for computing DFTs of length $2^r$ over $\mathbb{C}_p$, whose complexity is bounded by $O(\mathsf{I}(2^r p) + 2^r \mathsf{I}(p))$.*

**Proof.** Let $n := 2^r$, and suppose that we wish to compute the DFT of $a \in (\mathbb{C}_p \, 2^e)^n$. Using Bluestein's chirp transform (notation as in section 2.5), this reduces to computing a cyclic convolution of suitable $F \in (\mathbb{C}_p 2^e)[X]/(X^n - 1)$ and $G \in \mathbb{C}_p[X]/(X^n - 1)$. We assume that the $f_i$ and $g_i$ have been precomputed with $\rho_{f_i}, \rho_{g_i} \leqslant \varepsilon$.

We may regard $F' := 2^{p-e} F$ and $G' := 2^p G$ as cyclic polynomials with complex *integer* coefficients, i.e., as elements of $\mathbb{Z}[\mathrm{i}][X]/(X^n - 1)$. Write $F' = \sum_{i=0}^{n-1} F'_i X^i$ and $G' = \sum_{i=0}^{n-1} G'_i X^i$, where $F'_i, G'_i \in \mathbb{Z}[\mathrm{i}]$ with $|F'_i| \leqslant 2^p$ and $|G'_i| \leqslant 2^p$. Now we compute the *exact* product $H' := F' G' \in \mathbb{Z}[\mathrm{i}][X]/(X^n - 1)$ using Kronecker substitution. More precisely, we have $|H'_i| \leqslant 2^{2p+r}$, so it suffices to compute the cyclic integer product $H'(2^b) = F'(2^b) G'(2^b) \in (\mathbb{Z}/(2^{nb} - 1)\mathbb{Z})[\mathrm{i}]$, where $b := 2p + r + 2 = O(p)$. Then $H := H' 2^{e-2p}$ is the exact convolution of $F$ and $G$, and rounding $H$ to precision $p$ yields $F\mathtt{<dotast>}G \in (\mathbb{C}_p 2^{e+r})[X]/(X^n - 1)$ in the sense of Proposition 3.4. A final multiplication by $f_i$ yields the Fourier coefficients $\hat{a}_i \in \mathbb{C}_p 2^{e+r}$.

To establish tightness, observe that $1 + \rho_{F_i} \leqslant (1 + \rho_{a_i})(1 + \epsilon)^2$ and $\rho_{G_i} \leqslant \epsilon$, so Proposition 3.4 yields $1 + \rho_{(F\mathtt{<dotast>}G)_i} \leqslant (1 + \rho_a)(1 + \epsilon)^4$ where $\rho_a := \max_i \rho_{a_i}$; we conclude that $1 + \rho_{\hat{a}_i} \leqslant (1 + \rho_a)(1 + \epsilon)^6$. For $r \geqslant 3$, this means that the algorithm is tight; for $r \leqslant 2$, we may take $\mathcal{C}_r := \mathcal{B}^{\odot r}$.

For the complexity, observe that the product in $(\mathbb{Z}/(2^{nb} - 1)\mathbb{Z})[\mathrm{i}]$ reduces to three integer products of size $O(np)$. These have cost $O(\mathsf{I}(np))$, and the algorithm also performs $O(n)$ multiplications in $\mathbb{C}_p$, contributing the $O(n\mathsf{I}(p))$ term. □

**Remark 4.2.** A crucial observation is that, for suitable parameters, the DFT algorithm in Proposition 4.1 is actually faster than the conventional Cooley–Tukey algorithm of Corollary 3.9. For example, if we assume that $\mathsf{I}(m) = m(\log m)^{1+o(1)}$, then to compute a transform of length $n$ over $\mathbb{C}_p$ with $n \sim p$, the Cooley–Tukey approach has complexity $n^2(\log n)^{2+o(1)}$, whereas Proposition 4.1 yields $n^2(\log n)^{1+o(1)}$, an improvement by a factor of roughly $\log n$.

THEOREM 4.3. *For $n \to \infty$, we have*

$$\frac{\mathsf{I}(n)}{n \lg n} = O\left(\frac{\mathsf{I}(\lg^2 n)}{\lg^2 n \lg \lg n} + \frac{\mathsf{I}(\lg n)}{\lg n \lg \lg n} + 1\right). \tag{4.1}$$

**Proof.** We first reduce our integer product to a polynomial product using Kronecker segmentation (section 2.6). Splitting the two $n$-bit inputs into chunks of $b := \lg n$ bits, we need to compute a product of polynomials $u, v \in \mathbb{Z}[X]$ with non-negative $b$-bit coefficients and degrees less than $m := \lceil n/b \rceil = O(n/\lg n)$. The coefficients of $h := uv$ have $O(\lg n)$ bits, and we may deduce the desired integer product $h(2^b)$ in time $O(n)$.

Let $k := \lg(2m)$. To compute $uv$, we will use DFTs of length $2^k = O(n/\lg n)$ over $\mathbb{C}_p$, where $p := 2b + 2k + \lg k + 8 = O(\lg n)$. Zero-padding $u$ to obtain a sequence $(u_0, ..., u_{2^k-1}) \in (\mathbb{C}_p 2^b)^{2^k}$, and similarly for $v$, we compute the transforms $\hat{u}, \hat{v} \in (\mathbb{C}_p 2^{b+k})^{2^k}$ with respect to $\omega := e^{2\pi \mathrm{i}/2^k}$ as follows.

Let $r := \lg \lg n$ and $d := \lceil k/r \rceil = O(\lg n / \lg \lg n)$. Write $k = r_1 + \cdots + r_d$ with $r_i := r$ for $i \leqslant d - 1$ and $r_d := k - (d-1)r \leqslant r$. We use the algorithm $\mathcal{A} := \mathcal{A}_1 \odot \cdots \odot \mathcal{A}_d$ (see section 2.3), where for $1 \leqslant i \leqslant d - 1$ we take $\mathcal{A}_i$ to be the tight algorithm $\mathcal{C}_r$ for DFTs of length $2^r \asymp \lg n$ given by Proposition 4.1, and where $\mathcal{A}_d$ is $\mathcal{B}^{\odot r_d}$ as in Corollary 3.9. In other words, we split the $k$ usual radix-2 layers of the FFT into groups of $r$ layers, handling the transforms in each group with the Bluestein–Kronecker reduction, and then using ordinary Cooley–Tukey for the remaining $r_d$ layers.

We next compute the pointwise products $\hat{h}_i := \hat{u}_i \hat{v}_i \in \mathbb{C}_p 2^{2b+2k}$, and then apply an inverse transform $\mathcal{A}'$ defined analogously to $\mathcal{A}$. A final division by $2^k$ (which is really just an implicit adjustment of exponents) yields approximations $h_i \in \mathbb{C}_p 2^{2b+2k}$.

Since $\mathcal{A}$ and $\mathcal{A}'$ are tight by Propositions 3.8, 4.1 and Corollary 3.9, we have $1 + \rho_{\hat{u}_i} \leqslant (1 + \epsilon)^{3k-2}$, and similarly for $\hat{v}$. Thus $1 + \rho_{\hat{h}_i} \leqslant (1 + \epsilon)^{6k-3}$, so $1 + \rho_{h_i} \leqslant (1 + \epsilon)^{9k-5} \leqslant \exp(9k\epsilon) \leqslant \exp(2^{5+\lg k - p}) \leqslant 1 + 2^{6+\lg k - p}$ after the inverse transform (since $\exp x \leqslant 1 + 2x$ for $x \leqslant 1$). In particular, $\varepsilon_{h_i} = 2^{2b+2k} \rho_{h_i} \leqslant 2^{2b+2k+\lg k - p + 6} \leqslant 1/4$, so we obtain the exact value of $h_i$ by rounding to the nearest integer.

Now we analyse the complexity. Using Proposition 3.6, we first compute a table of roots $1, \omega, ..., \omega^{2^k-1}$ in time $O(2^k p \log p \log \log p) = O(n \lg n)$, and then extract the required twiddle factor tables in time $O(2^k k (p+k)) = O(n \lg n)$ (see section 2.3). For the Bluestein reductions, we may extract a table of $2^{r+1}$-th roots in time $O(2^k p) = O(n)$, and then rearrange them as required in time $O(2^r r (p+r)) = O(\lg^2 n \lg \lg n)$ (see section 2.5). These precomputations are then all repeated for the inverse transforms.

By Corollary 3.9, Proposition 4.1 and (2.4), each invocation of $\mathcal{A}$ (or $\mathcal{A}'$) has cost

$$O((d-1) 2^{k-r} (\mathsf{I}(2^r p) + 2^r \mathsf{I}(p)) + 2^{k-r_d} 2^{r_d} r_d \mathsf{I}(p) + (d-1) 2^k \mathsf{I}(p) + p\, 2^k k)$$
$$= O((d-1) 2^{k-r} \mathsf{I}(2^r p) + (d+r_d) 2^k \mathsf{I}(p) + p\, 2^k k)$$
$$= O\!\left( \frac{n}{\lg n \lg \lg n} \mathsf{I}(\lg^2 n) + \frac{n}{\lg \lg n} \mathsf{I}(\lg n) + n \lg n \right).$$

The cost of the $O(2^k)$ pointwise multiplications is subsumed within this bound. $\square$

It is now a straightforward matter to recover Fürer's bound.

THEOREM 4.4. *For some constant $K > 1$, we have*

$$\mathsf{I}(n) \;=\; O(n \lg n\, K^{\log^* n}).$$

**Proof.** Let $T(n) := \mathsf{I}(n)/(n \lg n)$ for $n \geqslant 2$. By Theorem 4.3, there exists $x_0 \geqslant 2$ and $C > 1$ such that

$$T(n) \;\leqslant\; C\, (T(\lg^2 n) + T(\lg n) + 1)$$

for all $n > x_0$. Let $\Phi(x) := 4 \log^2 x$ for $x \in \mathbb{R}$, $x > 1$. Increasing $x_0$ if necessary, we may assume that $\Phi(x) \leqslant x - 1$ for $x > x_0$, so that the function $\Phi^*(x) := \min\{j \in \mathbb{N} \colon \Phi^{\circ j}(x) \leqslant x_0\}$ is well-defined. Increasing $C$ if necessary, we may also assume that $T(n) \leqslant 3\, C$ for all $n \leqslant x_0$.

We prove by induction on $\Phi^*(n)$ that $T(n) \leqslant (3\, C)^{\Phi^*(n)+1}$ for all $n$. If $\Phi^*(n) = 0$, then $n \leqslant x_0$, so the bound holds. Now suppose that $\Phi^*(n) \geqslant 1$. Since $\lg^2 n \leqslant \Phi(n)$, we have $\Phi^*(\lg n) \leqslant \Phi^*(\lg^2 n) \leqslant \Phi^*(\Phi(n)) = \Phi^*(n) - 1$, so by induction $T(n) \leqslant C\, (3\, C)^{\Phi^*(n)} + C\, (3\, C)^{\Phi^*(n)} + C \leqslant (3\, C)^{\Phi^*(n)+1}$.

Finally, since $\Phi(\Phi(x)) \prec \log x$, we have $\Phi^*(x) \leqslant 2 \log^* x + O(1)$, so $T(n) = O(K^{\log^* n})$ for $K := (3\, C)^2$. $\square$

## 5. LOGARITHMICALLY SLOW RECURRENCE INEQUALITIES

This section is devoted to developing a framework for handling recurrence inequalities, similar to (4.1), that appear in subsequent sections.

Let $\Phi \colon (x_0, \infty) \to \mathbb{R}$ be a smooth increasing function, for some $x_0 \in \mathbb{R}$. We say that $\Phi^* \colon (x_0, \infty) \to \mathbb{R}^{\geqslant}$ is an *iterator* of $\Phi$ if $\Phi^*$ is increasing and if

$$\Phi^*(x) \;=\; \Phi^*(\Phi(x)) + 1 \tag{5.1}$$

for all sufficiently large $x$.

For instance, the standard iterated logarithm $\log^*$ defined in (1.2) is an iterator of $\log$. An analogous iterator may be defined for any smooth increasing function $\Phi \colon (x_0, \infty) \to \mathbb{R}$ for which there exists some $\sigma \geqslant x_0$ such that $\Phi(x) \leqslant x - 1$ for all $x > \sigma$. Indeed, in that case,

$$\Phi^*(x) \;:=\; \min\{k \in \mathbb{N} \colon \Phi^{\circ k}(x) \leqslant \sigma\}$$

is well-defined and satisfies (5.1) for all $x > \sigma$. It will sometimes be convenient to increase $x_0$ so that $\Phi(x) \leqslant x - 1$ is satisfied on the whole domain of $\Phi$.

We say that $\Phi$ is *logarithmically slow* if there exists an $\ell \in \mathbb{N}$ such that

$$(\log^{\circ \ell} \circ\, \Phi \circ \exp^{\circ \ell})(x) \;=\; \log x + O(1) \tag{5.2}$$

for $x \to \infty$. For example, the functions $\log(2\, x)$, $2 \log x$, $(\log x)^2$ and $(\log x)^{\log \log x}$ are logarithmically slow, with $\ell = 0, 1, 2, 3$ respectively.

LEMMA 5.1. *Let $\Phi\colon (x_0, \infty) \to \mathbb{R}$ be a logarithmically slow function. Then there exists $\sigma \geqslant x_0$ such that $\Phi(x) \leqslant x - 1$ for all $x > \sigma$. Consequently all logarithmically slow functions admit iterators.*

**Proof.** The case $\ell = 0$ is clear. For $\ell \geqslant 1$, let $\Psi := \log \circ \, \Phi \circ \exp$. By induction $\Psi(x) \leqslant x - 1$ for large $x$, so $\Phi(x) \leqslant \exp(\log x - 1) = x/e \leqslant x - 1$ for large $x$. □

In this paper, the main role played by logarithmically slow functions is to measure *size reduction* in multiplication algorithms. In other words, multiplication of objects of size $n$ will be reduced to multiplication of objects of size $n'$, where $n' \leqslant \Phi(n)$ for some logarithmically slow function $\Phi(x)$. The following result asserts that, from the point of view of iterators, such functions are more or less interchangeable with $\log x$.

LEMMA 5.2. *For any iterator $\Phi^*$ of a logarithmically slow function $\Phi$, we have*
$$\Phi^*(x) \;=\; \log^* x + O(1).$$

**Proof.** First consider the case where $\ell = 0$ in (5.2), i.e., assume that $|\Phi(x) - \log x| \leqslant C$ for some constant $C > 0$ and all $x > x_0$. Increasing $x_0$ and $C$ if necessary, we may assume that $\Phi^*(x) = \Phi^*(\Phi(x)) + 1$ for all $x > x_0$, and that $2\,\mathrm{e}^{2C} > x_0$.

We claim that
$$\frac{y}{2} \leqslant x \leqslant 2\,y \;\Longrightarrow\; \frac{\log y}{2} \leqslant \Phi(x) \leqslant 2 \log y \tag{5.3}$$
for all $y > 4\,e^{2C}$. Indeed, if $\frac{y}{2} \leqslant x \leqslant 2\,y$, then
$$\tfrac{1}{2} \log y \leqslant \log \tfrac{y}{2} - C \leqslant \Phi\bigl(\tfrac{y}{2}\bigr) \leqslant \Phi(x) \leqslant \Phi(2\,y) \leqslant \log(2\,y) + C \leqslant 2 \log y.$$

Now, given any $x > 4\,e^{2C}$, let $k := \min\{k \in \mathbb{N} : \log^{\circ k} x \leqslant 4\,\mathrm{e}^{2C}\}$, so $k \geqslant 1$. For any $j = 0, \ldots, k-1$ we have $\log^{\circ j} x > 4\,\mathrm{e}^{2C}$, so $k$-fold iteration of (5.3), starting with $y = x$, yields
$$\frac{\log^{\circ j} x}{2} \leqslant \Phi^{\circ j}(x) \leqslant 2 \log^{\circ j} x \quad (0 \leqslant j \leqslant k).$$

Moreover this shows that $\Phi^{\circ j}(x) > 2\,\mathrm{e}^{2C} > x_0$ for $0 \leqslant j < k$, so $\Phi^*(x) = \Phi^*(\Phi^{\circ k}(x)) + k$. Since $\Phi^{\circ k}(x) \leqslant 2 \log^{\circ k} x \leqslant 8\,\mathrm{e}^{2C}$ and $k = \log^* x + O(1)$, we obtain $\Phi^*(x) = \log^* x + O(1)$.

Now consider the general case $\ell \geqslant 0$. Let $\Psi := \log^{\circ \ell} \circ \Phi \circ \exp^{\circ \ell}$, so that $\Psi^* := \Phi^* \circ \exp^{\circ \ell}$ is an iterator of $\Psi$. By the above argument $\Psi^*(x) = \log^* x + O(1)$, and so $\Phi^*(x) = \Psi^*(\log^{\circ \ell} x) = \log^*(\log^{\circ \ell} x) + O(1) = \log^* x - \ell + O(1) = \log^* x + O(1)$. □

The next result, which generalises and refines the argument of Theorem 4.4, is our main tool for converting recurrence inequalities into actual asymptotic bounds for solutions. We state it in a slightly more general form than is necessary for the present paper, anticipating the more complicated situation that arises in [24].

PROPOSITION 5.3. *Let $K > 1$, $B \geqslant 0$ and $\ell \in \mathbb{N}$. Let $x_0 \geqslant \exp^{\circ \ell}(1)$, and let $\Phi\colon (x_0, \infty) \to \mathbb{R}$ be a logarithmically slow function such that $\Phi(x) \leqslant x - 1$ for all $x > x_0$. Then there exists a positive constant $C$ (depending on $x_0$, $\Phi$, $K$, $B$ and $\ell$) with the following property.*

*Let $\sigma \geqslant x_0$ and $L > 0$. Let $\mathcal{S} \subseteq \mathbb{R}$, and let $T\colon \mathcal{S} \to \mathbb{R}^{\geqslant}$ be any function satisfying the following recurrence. First, $T(y) \leqslant L$ for all $y \in \mathcal{S}$, $y \leqslant \sigma$. Second, for all $y \in \mathcal{S}$, $y > \sigma$, there exist $y_1, \ldots, y_d \in \mathcal{S}$ with $y_i \leqslant \Phi(y)$, and weights $\gamma_1, \ldots, \gamma_d \geqslant 0$ with $\sum_i \gamma_i = 1$, such that*
$$T(y) \;\leqslant\; K\left(1 + \frac{B}{\log^{\circ \ell} y}\right) \sum_{i=1}^{d} \gamma_i T(y_i) + L.$$

*Then we have $T(y) \leqslant C L K^{\log^* y - \log^* \sigma}$ for all $y \in \mathcal{S}$, $y > \sigma$.*

**Proof.** Let $\sigma$, $L$, $\mathcal{S}$ and $T(x)$ be as above. Define $\Phi^*_\sigma(x) := \min\{k \in \mathbb{N} : \Phi^{\circ k}(x) \leqslant \sigma\}$ for $x > x_0$. We claim that there exists $r \in \mathbb{N}$, depending only on $x_0$ and $\Phi$, such that
$$\Phi^*_\sigma(x) \;\leqslant\; \log^* x - \log^* \sigma + r \tag{5.4}$$

for all $x > \sigma$. Indeed, let $\Phi^*(x) := \min\{j \in \mathbb{N} : \Phi^{\circ j}(x) \leqslant x_0\}$. First suppose $\sigma > x_0$, so that $\Phi^*(\sigma) \geqslant 1$. For any $x > \sigma$, we have $\Phi^{\circ(\Phi_\sigma^*(x)-1)}(x) > \sigma$, so

$$\Phi^{\circ(\Phi_\sigma^*(x)-1+\Phi^*(\sigma)-1)}(x) \geqslant \Phi^{\circ(\Phi^*(\sigma)-1)}(\sigma) > x_0,$$

and hence $\Phi^*(x) > \Phi_\sigma^*(x) + \Phi^*(\sigma) - 2$. This last inequality also clearly holds if $\sigma = x_0$ (since $0 > -2$). By Lemma 5.2 we obtain $\Phi_\sigma^*(x) \leqslant \Phi^*(x) - \Phi^*(\sigma) + O(1) = \log^* x - \log^* \sigma + O(1)$.

Define a sequence of real numbers $E_1, E_2, \ldots$ by the formula

$$E_j := \begin{cases} 1 + B & \text{if } j \leqslant r + \ell, \\ 1 + B/\exp^{\circ(j-r-\ell-1)}(1) & \text{if } j > r + \ell. \end{cases}$$

We claim that

$$1 + B/\log^{\circ \ell} x \;\leqslant\; E_{\Phi_\sigma^*(x)} \tag{5.5}$$

for all $x > \sigma$. Indeed, let $j := \Phi_\sigma^*(x)$. If $j \leqslant r + \ell$ then (5.5) holds as $x > \sigma \geqslant x_0 \geqslant \exp^{\circ \ell}(1)$. If $j > r + \ell$ then $\log^* x \geqslant j - r$ by (5.4), so $x \geqslant \exp^{\circ(j-r-1)}(1)$ and hence $\log^{\circ \ell} x \geqslant \exp^{\circ(j-r-\ell-1)}(1)$.

Now let $y \in \mathcal{S}$. We will prove by induction on $j := \Phi_\sigma^*(y)$ that

$$T(y) \;\leqslant\; E_1 \cdots E_j L \, (K^j + \cdots + K + 1)$$

for all $y > x_0$. The base case $j := 0$, i.e., $y \leqslant \sigma$, holds by assumption. Now assume that $j \geqslant 1$, so $y > \sigma$. By hypothesis there exist $y_1, \ldots, y_d \in \mathcal{S}$, $y_i \leqslant \Phi(y)$, and $\gamma_1, \ldots, \gamma_d \geqslant 0$ with $\sum_i \gamma_i = 1$, such that

$$T(y) \;\leqslant\; K E_j \sum_i \gamma_i T(y_i) + L.$$

Since $\Phi_\sigma^*(y_i) \leqslant \Phi_\sigma^*(\Phi(y)) = \Phi_\sigma^*(y) - 1$, we obtain

$$\begin{aligned} T(y) &\leqslant K E_j \sum_i \gamma_i (E_1 \cdots E_{j-1} L \, (K^{j-1} + \cdots + K + 1)) + L \\ &= E_1 \cdots E_j L \, (K^j + \cdots + K^2 + K) + L \\ &\leqslant E_1 \cdots E_j L \, (K^j + \cdots + K^2 + K + 1). \end{aligned}$$

Finally, the infinite product

$$E := \prod_{j \geqslant 1} E_j \leqslant (1+B)^{r+\ell} \prod_{k \geqslant 0} \left(1 + \frac{B}{\exp^{\circ k}(1)}\right)$$

certainly converges, so we have $T(y) \leqslant E L K^{j+1}/(K-1)$ for $y > x_0$. Setting $C := E K^{r+1}/(K-1)$, by (5.4) we obtain $T(y) \leqslant C L K^{\log^* y - \log^* \sigma}$ for all $y > \sigma$. □

## 6. EVEN FASTER MULTIPLICATION

In this section, we present an optimised version of the new integer multiplication algorithm. The basic outline is the same as in section 4, but our goal is now to minimise the "expansion factor" at each recursion level. The necessary modifications may be summarised as follows.

- Since Bluestein's chirp transform reduces a DFT to a complex cyclic convolution, we take the basic recursive problem to be complex cyclic integer convolution, i.e., multiplication in $(\mathbb{Z}/(2^n - 1)\mathbb{Z})[i]$, rather than ordinary integer multiplication.

- In multiplications involving one fixed operand, we reuse the transform of the fixed operand.

- In a convolution of length $n$ with input coefficients of bit size $b$, the size of the output coefficients is $2b + O(\lg n)$, so the ratio of output to input size is $2 + O((\lg n)/b)$. We increase $b$ from $\lg n$ to $(\lg n)^2$, so as to reduce the inflation ratio from $O(1)$ to $2 + O(1/\lg n)$.

- We increase the "short transform length" from $\lg n$ to $(\lg n)^{\lg \lg n + O(1)}$. The complexity then becomes dominated by the Bluestein–Kronecker multiplications, while the contribution from ordinary arithmetic in $\mathbb{C}_p$ becomes asymptotically negligible. (As noted in section 1, this is precisely the opposite of what occurs in Fürer's algorithm.)

We begin with a technical preliminary. To perform multiplication in $(\mathbb{Z}/(2^n-1)\mathbb{Z})[i]$ efficiently using FFT multiplication, we need $n$ to be divisible by a high power of two. We say that an integer $n \geqslant 3$ is *admissible* if $2^{\kappa(n)} \mid n$, where $\kappa(n) := \lg n - \lg(\lg^2 n) + 1$ (note that $0 \leqslant \kappa(n) \leqslant \lg n$ for all $n \geqslant 3$). We will need a function that rounds a given $n$ up to an admissible integer. For this purpose we define $\alpha(n) := \lceil n/2^{\kappa(n)} \rceil 2^{\kappa(n)}$ for $n \geqslant 3$. Note that $\alpha(n)$ may be computed in time $O(\lg n)$.

LEMMA 6.1. *Let $n \geqslant 3$. Then $\alpha(n)$ is admissible and*

$$n \leqslant \alpha(n) \leqslant n + \frac{4n}{\lg^2 n}. \tag{6.1}$$

**Proof.** We have $n \leqslant \alpha(n) \leqslant n + 2^{\kappa(n)}$, which implies (6.1). Since $n/2^{\kappa(n)} \leqslant 2^{\lg n - \kappa(n)}$ and $\kappa(n) \leqslant \lg n$, we have $\lceil n/2^{\kappa(n)} \rceil \leqslant 2^{\lg n - \kappa(n)}$ and thus $\alpha(n) \leqslant 2^{\lg n}$, i.e., $\lg \alpha(n) = \lg n$. In particular $\kappa(\alpha(n)) = \kappa(n)$, so $\alpha(n)$ is admissible. (In fact, one easily checks that $\alpha(n)$ is the *smallest* admissible integer $\geqslant n$). □

**Remark 6.2.** It is actually possible to drop the requirement that $n$ be divisible by a high power of two, by using the Crandall–Fagin method (see section 9). We prefer to avoid this approach in this section, as it adds an unnecessary layer of complexity to the presentation.

Now let $n$ be admissible, and consider the problem of computing $t \geqslant 1$ products $u_1 v, \ldots, u_t v$ with $u_1, \ldots, u_t, v \in (\mathbb{Z}/(2^n-1)\mathbb{Z})[i]$, i.e., $t$ products with one fixed operand. Denote the cost of this operation by $\mathsf{C}_t(n)$. Our algorithm for this problem will perform $t+1$ forward DFTs and $t$ inverse DFTs, so it is convenient to introduce the normalisation

$$\mathsf{C}(n) := \sup_{t \geqslant 1} \frac{\mathsf{C}_t(n)}{2t+1}.$$

This is well-defined since clearly $\mathsf{C}_t(n) \leqslant t\,\mathsf{C}_1(n)$. Roughly speaking, $\mathsf{C}(n)$ may be thought of as the notional cost of a single DFT.

The problem of multiplying $k$-bit integers may be reduced to the above problem by using zero-padding, i.e., by taking $n := \alpha(2k+1)$ and $t := 1$. Since $\alpha(2k+1) = O(k)$ and $\mathsf{C}_1(n) \leqslant 3\,\mathsf{C}(n)$, we obtain $\mathsf{I}(k) \leqslant 3\,\mathsf{C}(O(k)) + O(k)$. Thus it suffices to obtain a good bound for $\mathsf{C}(n)$.

The recursive step in the main multiplication algorithm involves computing "short" DFTs via the Bluestein–Kronecker device. As pointed out in section 2.5, this leads to a cyclic convolution with one fixed operand. To take advantage of the fixed operand, let $\mathsf{B}_{p,t}(2^r)$ denote the cost of computing $t$ independent DFTs of length $2^r$ over $\mathbb{C}_p$, and let $\mathsf{B}_p(2^r) := \sup_{t \geqslant 1} \mathsf{B}_{p,t}(2^r)/(2t+1)$. Then we have the following refinement of Proposition 4.1. As usual we assume that the necessary Bluestein root table has been precomputed.

PROPOSITION 6.3. *Let $r \geqslant 3$, and assume that $2^r$ divides $n' := \alpha((2p+r+2)\,2^r)$. Then there exists a tight algorithm $\mathcal{C}'_r$ for computing DFTs of length $2^r$ over $\mathbb{C}_p$, with*

$$\mathsf{B}_p(2^r) \leqslant \mathsf{C}(n') + O(2^r\,\mathsf{I}(p)).$$

**Proof.** We use the same notation and algorithm as in the proof of Proposition 4.1, except that in the Kronecker substitution we take $b := n'/2^r \geqslant 2p+r+2$, so that the resulting integer multiplication takes place in $(\mathbb{Z}/(2^{n'}-1)\mathbb{Z})[i]$. The proof of tightness is identical to that of Proposition 4.1 (this is where we use the assumption $r \geqslant 3$). For the complexity bound, note that $n'$ is admissible by construction, so for any $t \geqslant 1$ we have $\mathsf{B}_{p,t}(2^r) \leqslant \mathsf{C}_t(n') + O(t\,2^r\,\mathsf{I}(p))$. Here we have used the fact that $G'$ is fixed over all these multiplications. Dividing by $2t+1$ and taking suprema over $t \geqslant 1$ yields the result. □

The next result gives the main recurrence satisfied by $\mathsf{C}(n)$ (compare with Theorem 4.3).

THEOREM 6.4. *There exists $x_0 \geqslant 3$ and a logarithmically slow function $\Phi: (x_0, \infty) \to \mathbb{R}$ with the following property. For all admissible $n > x_0$, there exists an admissible $n' \leqslant \Phi(n)$ such that*

$$\frac{\mathsf{C}(n)}{n \lg n} \leqslant \left(8 + O\!\left(\frac{1}{\lg \lg n}\right)\right) \frac{\mathsf{C}(n')}{n' \lg n'} + O(1). \tag{6.2}$$

**Proof.** Let $n$ be admissible and sufficiently large, and consider the problem of computing $t \geqslant 1$ products $u_1 v, \ldots, u_t v$, for $u_1, \ldots, u_t, v \in (\mathbb{Z}/(2^n-1)\mathbb{Z})[\mathrm{i}]$. Let $k := \kappa(n) \sim \lg n$, so that $2^k | n$, and let $b := n/2^k \asymp \lg^2 n$.

We cut the inputs into $2^k$ chunks of size $b$, i.e., if $w$ is one of the $t+1$ inputs, we write $w = w_0 + w_1 2^b + \cdots + w_{2^k-1} 2^{(2^k-1)b}$, where $w_i \in \mathbb{Z}[\mathrm{i}]$, and where the real and imaginary parts of $w_i$ have absolute value at most $2^b$. Thus $|w_i| \leqslant \sqrt{2} \cdot 2^b < 2^{b+1}$, and for any $p \geqslant b+1$ we may encode $w$ as a polynomial $W \in (\mathbb{C}_p 2^{b+1})[X]/(X^{2^k} - 1)$.

We will multiply the desired (cyclic) polynomials by using DFTs of length $2^k$ over $\mathbb{C}_p$ where $p := 2b + 2k + \lg k + 10 = O(\lg^2 n)$. We construct the DFTs in a similar way to section 4. Let $r := (\lg \lg n)^2$ and $d := \lceil k/r \rceil = O(\lg n/(\lg \lg n)^2)$. Write $k = r_1 + \cdots + r_d$ with $r_i := r$ for $i \leqslant d-1$ and $r_d := k - (d-1)r \leqslant r$. We use the tight algorithm $\mathcal{A} := \mathcal{A}_1 \odot \cdots \odot \mathcal{A}_d$, where for $1 \leqslant i \leqslant d-1$ we take $\mathcal{A}_i$ to be the tight algorithm $\mathcal{C}'_r$ for DFTs of length $2^r$ given by Proposition 6.3, and where $\mathcal{A}_d$ is $\mathcal{B}^{\odot r_d}$ as in Corollary 3.9. Thus, for the first $d-1$ groups of $r$ layers, we use Bluestein–Kronecker to reduce to complex integer convolution of size $n' := \alpha((2p + r + 2)2^r)$, and the remaining layers are handled using ordinary Cooley–Tukey. We write $\mathcal{A}'$ for the analogous inverse transform.

To check the hypothesis of Proposition 6.3, we observe that $2^r | n'$ for sufficiently large $n$, as $n'$ is divisible by $2^{k'}$ where $k' := \lg n' - \lg(\lg^2 n') + 1$, and

$$2^{k'} \asymp \frac{n'}{\lg^2 n'} \asymp \frac{(2p+r+2)2^r}{\lg^2((2p+r+2)2^r)} \asymp \frac{b\,2^r}{(\lg b + r)^2} \asymp \frac{(\lg n)^2}{(\lg \lg n)^4} 2^r \succ 2^r.$$

Denote by $\mathsf{D}$ the cost of a single invocation of $\mathcal{A}$ (or $\mathcal{A}'$). By Corollary 3.9 and (2.4), we have

$$\mathsf{D} \;\leqslant\; (d-1)\mathsf{B}_{p,2^{k-r}}(2^r) + O(2^{k-r_d} 2^{r_d} r_d \mathsf{I}(p)) + O(d\, 2^k \mathsf{I}(p)) + O(2^k k b).$$

The last term is the rearrangement cost, and simplifies to $O(n \lg n)$. The second term covers the invocations of $\mathcal{A}_d$, and simplifies to $O(r\, 2^k \mathsf{I}(p))$, so is absorbed by the $d\, 2^k \mathsf{I}(p)$ term. The first term covers the invocations of $\mathcal{C}'_r$. By definition $\mathsf{B}_{p,2^{k-r}}(2^r) \leqslant (2 \cdot 2^{k-r} + 1)\mathsf{B}_p(2^r)$, and since $2^{k-r} \succ \lg \lg n$, Proposition 6.3 yields

$$\mathsf{B}_{p,2^{k-r}}(2^r) \;\leqslant\; (2 + O(1/\lg \lg n))\, 2^{k-r}\, \mathsf{C}(n') + O(2^k \mathsf{I}(p)).$$

Thus

$$\mathsf{D} \;\leqslant\; (2 + O(1/\lg \lg n))\, d\, 2^{k-r}\, \mathsf{C}(n') + O(d\, 2^k \mathsf{I}(p)) + O(n \lg n).$$

We will use Schönhage–Strassen's algorithm for fixed point multiplications in $\mathbb{C}_p$. Since $p = O(\lg^2 n)$, we may take $\mathsf{I}(p) = O(\lg^2 n \lg \lg n \lg \lg \lg n)$. Thus the $d\, 2^k \mathsf{I}(p)$ term becomes

$$O\!\left(\frac{\lg n}{(\lg \lg n)^2} \frac{n}{\lg^2 n} \lg^2 n \lg \lg n \lg \lg \lg n\right) = O\!\left(n \lg n \frac{\lg \lg \lg n}{\lg \lg n}\right) = O(n \lg n).$$

(We could of course use our algorithm recursively for these multiplications; however, it turns out that Schönhage–Strassen is fast enough, and leads to simpler recurrences. In fact, the algorithm asymptotically spends more time rearranging data than multiplying in $\mathbb{C}_p$!)

Since $(2p + r + 2) 2^r = (4b + O(\lg n)) 2^r = (4 + O(1/\lg \lg n)) b\, 2^r$, and since $\lg(b\, 2^r) = r + O(\lg \lg n) = (1 + O(1/\lg \lg n)) r \succ \lg \lg n$, by Lemma 6.1 we have

$$n' = (4 + O(1/\lg \lg n)) b\, 2^r,$$
$$\lg n' = (1 + O(1/\lg \lg n)) r.$$

We also have $k = \lg n + O(\lg \lg n)$ and $d = k/r + O(1)$, so

$$\lg n = (1 + O(1/\lg \lg n)) k,$$
$$d = (1 + O(1/\lg \lg n)) k/r.$$

Thus

$$d\, 2^{k-r} = \frac{4(2^k b)\, d}{(4b\, 2^r)} = \left(4 + O\!\left(\frac{1}{\lg \lg n}\right)\right) \frac{n \lg n}{n' \lg n'},$$

and consequently

$$\mathsf{D} \;\leqslant\; \left(8 + O\!\left(\frac{1}{\lg \lg n}\right)\right) \frac{n \lg n}{n' \lg n'} \mathsf{C}(n') + O(n \lg n).$$

To compute the desired $t$ products, we must execute $t+1$ forward transforms and $t$ inverse transforms. For each product, we must also perform $O(2^k)$ pointwise multiplications in $\mathbb{C}_p$, at cost $O(2^k \mathsf{I}(p)) = O(n \lg n)$. As in the proof of Theorem 4.3, the cost of all necessary root table precomputations is also bounded by $O(2^k \mathsf{I}(p)) = O(n \lg n)$. Thus we obtain

$$\mathsf{C}_t(n) \;\leqslant\; (2\,t+1)\,\mathsf{D} + O(t\,n \lg n).$$

Dividing by $(2\,t+1)\,n \lg n$ and taking suprema yields the bound (6.2).

The error analysis is almost identical to the proof of Theorem 4.3, the only difference being that $b$ is replaced by $b+1$. Denoting one of the $t$ products by $h \in (\mathbb{C}_p\,2^{2b+2k+2})[X]/(X^{2^k}-1)$, we have $\rho_{h_i} \leqslant 2^{6+\lg k-p}$ exactly as in Theorem 4.3. Thus $\varepsilon_{h_i} \leqslant 2^{2b+2k+\lg k-p+8} \leqslant 1/4$, and again we obtain $h_i$ by rounding to the nearest integer.

Finally we show how to define $\Phi(x)$. We already observed that $\lg n' \sim r \sim (\lg \lg n)^2$. Thus there exists a constant $C > 0$ such that $\log \log \log n' \leqslant \log \log \log \log n + C$ for large $n$, so we may take $\Phi(x) := \exp^{\circ 3}(\log^{\circ 4} x + C)$. $\square$

Now we may prove the main theorem announced in the introduction.

**Proof of Theorem 1.1.** Let $x_0$ and $\Phi(x)$ be as in Theorem 6.4. Increasing $x_0$ if necessary, by Lemma 5.1 we may assume that $\Phi(x) \leqslant x-1$ for $x > x_0$, and that $x_0 \geqslant \exp(\exp(1))$.

Let $T(n) := \mathsf{C}(n)/(n \lg n)$ for admissible $n \geqslant 3$. By the theorem, there exist constants $B, L > 0$ such that for all admissible $n > x_0$, there exists an admissible $n' \leqslant \Phi(n)$ with

$$T(n) \;\leqslant\; 8\left(1 + \frac{B}{\log \log n}\right) T(n') + L.$$

Increasing $L$ if necessary, we may also assume that $T(n) \leqslant L$ for all admissible $n \leqslant x_0$. Taking $\mathcal{S}$ to be the set of admissible integers, we apply Proposition 5.3 with $K := 8$, $\sigma := x_0$, $\ell := 2$, and for each admissible $n > x_0$ setting $d := 1$, $\gamma_1 := 1$, $y := n$ and $y_1 := n'$ as above. We conclude that $T(n) = O(8^{\log^* n})$, and hence $\mathsf{C}(n) = O(n \lg n\, 8^{\log^* n})$ as $n$ runs over admissible integers. We already pointed out that $\mathsf{I}(k) \leqslant 3\,\mathsf{C}(O(k)) + O(k)$. $\square$

## 7. An optimised variant of Fürer's algorithm

As pointed out in the introduction, Fürer proved that $\mathsf{I}(n) = O(n \log n\, K^{\log^* n})$ for some $K > 1$, but did not give an explicit bound for $K$. In this section we sketch an argument showing that one may achieve $K = 16$ in Fürer's algorithm, by reusing tools from previous sections, especially section 6.

At the core of Fürer's algorithm is the ring $R = \mathbb{C}[X]/(X^{2^{r-1}}+1)$, which contains the principal $2^r$-th root of unity $X$. Note that $R$ is a direct sum of $2^{r-1}$ copies of $\mathbb{C}$, and hence not a field (for $r \geqslant 2$). A crucial observation is that $X$ is a "fast" root of unity, in the sense that multiplication by $X$ and its powers can be achieved in linear time, as in Schönhage–Strassen's algorithm. For any $k > r$, we need to construct a $2^{k-r}$-th root $\omega$ of $X$, which is itself a $2^k$-th principal root of unity. We recall Fürer's construction of $\omega$ as follows.

**Lemma 7.1.** *With $R$ as above, let $\varrho = \exp \frac{2\pi i}{2^k}$ and $\sigma = \exp \frac{2\pi i}{2^r}$. Then*

$$\omega \;:=\; \sum_{i=0}^{2^{r-1}-1} \varrho^{2i+1}\, \frac{\prod_{j \neq i}(X - \sigma^{2j+1})}{\prod_{j \neq i}(\sigma^{2i+1} - \sigma^{2j+1})} \quad \in R$$

*is a principal $2^k$-th root of unity with $\omega^{2^{k-r}} = X$. The coefficients of $\omega$ have absolute value $\leqslant 1$.*

**Proof.** See [19, Section 4]. $\square$

As our basic recursive problem, we will consider multiplication in $(\mathbb{Z}/(2^n+1)\,\mathbb{Z})[i]$, where $n$ is divisible by a high power of two. We will refer to the last property as "admissibility", but we will not define it precisely. We write $\mathsf{C}_t(n)$ for the cost of $t \geqslant 1$ such products with one fixed argument, and $\mathsf{C}(n) := \sup_{t \geqslant 1} \mathsf{C}_t(n)/(2\,t+1)$ for the normalised cost, exactly as in section 6.

Fürer worked with $\mathbb{Z}/(2^n+1)\mathbb{Z}$ rather than $(\mathbb{Z}/(2^n+1)\mathbb{Z})[i]$, but, since we are interested in constant factors, and since the recursive multiplication step involves multiplication of complex quantities, it simplifies the exposition to work systematically with complexified objects everywhere.

For suitable parameters $r$ and $k$, we will encode elements of $(\mathbb{Z}/(2^n+1)\mathbb{Z})[i]$ as (nega)cyclic polynomials in $R[Y]/(Y^{2^k}+1)$, where $R := \mathbb{C}[X]/(X^{2^{r-1}}+1)$ as above. We choose the parameters later; for now we require only that $2^{k+r-2}$ divides $n$ and that $b := n/2^{k+r-2} \geqslant \lg n$ (so that the coefficients are not too small).

The encoding proceeds as follows. Given $a \in \mathbb{Z}/(2^n+1)\mathbb{Z}$, we split $a$ into $2^k$ parts $a_0, ..., a_{2^k-1}$ of $n/2^k$ bits. Each $a_i$ is cut into $2^{r-2}$ even smaller pieces $a_{i,0}, ..., a_{i,2^{r-2}-1}$ of $b$ bits. Then $a$ is encoded as

$$\tilde{a} := \sum_{i=0}^{2^k-1} \sum_{j=0}^{2^{r-2}-1} a_{i,j} X^j Y^i,$$

and an element $u = x + y\,\mathrm{i} \in (\mathbb{Z}/(2^n+1)\mathbb{Z})[i]$ is encoded as $\tilde{u} := \tilde{x} + \tilde{y}\,\mathrm{i}$. (Notice that the coefficients of $X^j$ are zero for $2^{r-2} \leqslant j < 2^{r-1}$; this zero-padding is the price Fürer pays for introducing artificial roots of unity.)

We represent complex coefficients by elements of $\mathbb{C}_p \, 2^e$ for a suitable precision parameter $p$. The exponent $e$ varies during the algorithm, as explained in [19]; nevertheless, additions and subtractions only occur for numbers with the same exponent, as in the algorithms from sections 4 and 6.

Given $u, v \in (\mathbb{Z}/(2^n+1)\mathbb{Z})[i]$, to successfully recover the product $uv$ from the polynomial product $\tilde{u}\tilde{v} \in R[Y]/(Y^{2^k}+1)$, we must choose $p \geqslant 2b + k + r + h$, where $h$ is an allowance for numerical error. Certainly $r \leqslant k \leqslant \lg n$, and, as shown by Fürer, we may also take $h = O(\lg n)$ (an analogous conclusion is reached in sections 4 and 6). Thus we may assume that $p = 2b + O(\lg n)$.

We must now show how to compute a product $\tilde{u}\tilde{v}$, for $\tilde{u}, \tilde{v} \in R[Y]/(Y^{2^k}+1)$. Fürer handles these types of multiplications using "half-DFTs", i.e., DFTs that evaluate at odd powers of $\eta$, where $\eta \in R$ is a principal $2^{k+1}$-th root of unity such that $\eta^{2^{k+1-r}} = X$ (Lemma 7.1). To keep terminology and notation consistent with previous sections, we prefer to make the substitution $U(X, Y) := \tilde{u}(X, \eta Y)$, i.e., writing $\tilde{u} = \sum_{i=0}^{2^k-1} \tilde{u}_i(X) Y^i$, we put $U := \sum_i (\tilde{u}_i \eta^i) Y^i$, and similarly for $\tilde{v}$ and $V$. This reduces the problem to computing the product $UV$ in $R[Y]/(Y^{2^k}-1)$. The change of variable imposes a cost of $O(2^k \mathsf{m}_R)$, where $\mathsf{m}_R$ is the cost of a multiplication in $R$.

So now consider a product $UV$, where $U, V \in R[Y]/(Y^{2^k}-1)$. Let $\omega := \eta^2$, so that $\omega^{2^{k-r}} = X$. Let $d := \lceil k/r \rceil$, and write $k = r_1 + \cdots + r_d$ with $r_i := r$ for $i \leqslant d-1$ and $r_d := k - (d-1)r \leqslant r$. For each $i$, let $\mathcal{A}_i$ be the algorithm for DFTs of length $2^{r_i}$ that applies the usual Cooley–Tukey method, taking advantage of the fast $2^{r_i}$-th root of unity $X^{2^{r-r_i}}$. The complexity of $\mathcal{A}_i$ is $O(2^{r_i+r} r_i p)$, since it performs $O(2^{r_i} r_i)$ linear-time operations on objects of bit size $O(2^r p)$. Let $\mathsf{D}$ be the complexity of the algorithm $\mathcal{A} := \mathcal{A}_1 \odot \cdots \odot \mathcal{A}_d$ for DFTs of length $2^k$ over $R$. Then (2.4) yields

$$\mathsf{D} \;\leqslant\; O\!\left(\sum_{i=1}^d 2^{k-r_i} 2^{r_i+r} r_i p\right) + \left\lceil \frac{k}{r} \right\rceil 2^k \mathsf{m}_R + O(n \lg n),$$

The first term is bounded by $O(d\, 2^k\, 2^r\, r\, p) = O((2^{k+r} p)\, k) = O(n \lg n)$, since $p = O(b)$.

Let us now consider the second term $\lceil k/r \rceil 2^k \mathsf{m}_R$, which describes the cost of the twiddle factor multiplications. This term turns out to be the dominant one. Both Kronecker substitution and FFT multiplication may be considered for multiplication in $R$, but it turns out that Kronecker substitution is faster (a similar phenomenon was noted in Remark 4.2). So we reduce multiplication in $R$ to multiplication in $(\mathbb{Z}/(2^{n'}+1)\mathbb{Z})[i]$ where $n' \geqslant 2^{r-1}(2p+r+2)$ is admissible and divisible by $2^{r-1}$. For any reasonable definition of admissibility we then have $n' = (1+o(1))\, 2^r p$, provided that $r$ is somewhat smaller than $p$. (In the interests of brevity, we will not specify the $o(1)$ terms for the remainder of the argument. They can all be controlled along the lines of section 6.) Most of the twiddle factors are reused many times, so we will assume that $\mathsf{m}_R = (2+o(1))\,\mathsf{C}(n')$, where the factor 2 counts the two (rather than three) DFTs needed for each multiplication of size $n'$. The term of interest then becomes

$$\left\lceil \frac{k}{r} \right\rceil 2^k \mathsf{m}_R \;=\; (2+o(1))\, \frac{r+\lg p}{r}\, \frac{2^{k+r} p\, k}{n' \lg n'}\, \mathsf{C}(n').$$

Since $p = 2b + O(\lg n) = \left(2 + O\left(\frac{\lg n}{b}\right)\right) b$ and $2^{k+r} b = 4n$, this yields

$$\mathsf{D} \;\leqslant\; (16 + o(1)) \left(1 + O\left(\frac{\lg n}{b}\right)\right) \frac{r + \lg p}{r} \frac{n \lg n}{n' \lg n'} \mathsf{C}(n') + O(n \lg n).$$

To minimise the leading constant, we must choose $b$ to grow faster than $\lg n$, and $r$ to grow faster than $\lg p$. For example, taking $r := (\lg \lg n)^2$ and $k := \lg n - r - \lg(\lg^2 n)$ leads to $b = 4n / 2^{k+r} \asymp \lg^2 n$ and $\lg p \asymp \lg b \asymp \lg \lg n$. The function mapping $n$ to $n'$ is then bounded by a logarithmically slow function, and a similar argument to section 6 shows that $\mathsf{I}(n) = O(n \log n \, 16^{\log^* n})$.

## 8. FAST MULTIPLICATION USING MODULAR ARITHMETIC

Shortly after Fürer's algorithm appeared, De et al [15] presented a variant based on modular arithmetic that also achieves the complexity bound $\mathsf{I}(n) = O(n \log n \, K^{\log^* n})$ for some $K > 1$. Roughly speaking, they replace the coefficient ring $\mathbb{C}$ with the field $\mathbb{Q}_p$ of $p$-adic numbers, for a suitable prime $p$. In this context, working to "finite precision" means performing computations in $\mathbb{Z}/p^\lambda \mathbb{Z}$, where $\lambda \geqslant 1$ is a precision parameter.

The main advantage of this approach is that the error analysis becomes trivial; indeed $\mathbb{Z}/p^\lambda \mathbb{Z}$ is a ring (unlike our $\mathbb{C}_p$), and arithmetic operations never lead to precision loss (unless one divides by $p$, which never happens in these algorithms). The main disadvantage is that there are certain technical difficulties associated with finding an appropriate $p$; this is discussed in section 8.2 below.

The aim of this section is to sketch an analogue of the algorithm of section 6 that achieves $\mathsf{I}(n) = O(n \log n \, 8^{\log^* n})$ using modular arithmetic instead of $\mathbb{C}$. We assume familiarity with $p$-adic numbers, referring the reader to [22] for an elementary introduction.

### 8.1. Sketch of the algorithm

For the basic problem, we take multiplication in $\mathbb{Z}/(2^n - 1)\mathbb{Z}$, where $n$ is admissible (in the sense of section 6) and where one of the arguments is fixed over $t \geqslant 1$ multiplications. As before, we take $k := \kappa(n)$, and cut the inputs into chunks of $b := n/2^k = O(\lg^2 n)$ bits. Thus we reduce to multiplying polynomials in $\mathbb{Z}[X]/(X^{2^k} - 1)$ with coefficients of at most $b$ bits. The coefficients of the product have at most $2b + k$ bits.

Let $p$ be a prime such that $p \equiv 1 \pmod{2^k}$, so that $\mathbb{Q}_p$ contains a primitive $2^k$-th root of unity $\omega$. The problem of finding such $p$ and $\omega$ is discussed in the next section; for now we assume only that $\lg p = O(\lg n)$. We may then embed the multiplication problem into $\mathbb{Q}_p[X]/(X^{2^k} - 1)$, and use DFTs with respect to $\omega$ to compute the product. On a Turing machine, we cannot represent elements of $\mathbb{Q}_p$ exactly, so we perform all computations in $\mathbb{Z}/p^\lambda \mathbb{Z}$ where

$$\lambda \;:=\; \left\lceil \frac{2b + k}{(\lg p) - 1} \right\rceil.$$

This choice ensures that $\lg(p^\lambda) \geqslant 2b + k$, so knowledge of the product in $(\mathbb{Z}/p^\lambda \mathbb{Z})[X]/(X^{2^k} - 1)$ determines it unambiguously in $\mathbb{Z}[X]/(X^{2^k} - 1)$.

To compute each DFT, we first use the Cooley–Tukey algorithm to decompose it into "short transforms" of length $2^r$, where $r := (\lg \lg n)^2$. (As in section 6, there are also residual transforms of length $2^{r_d}$ for some $r_d \leqslant r$, whose contribution to the complexity is negligible.) Multiplications in $\mathbb{Z}/p^\lambda \mathbb{Z}$, such as the multiplications by twiddle factors, are handled using Schönhage–Strassen's algorithm, with the divisions by $p^\lambda$ being reduced to multiplication via Newton's method. We then use Bluestein's algorithm to convert each short transform to a cyclic convolution of length $2^r$ over $\mathbb{Z}/p^\lambda \mathbb{Z}$, and apply Kronecker substitution to convert this to multiplication in $\mathbb{Z}/(2^{n'} - 1)\mathbb{Z}$, where $n'$ is the smallest admissible integer exceeding $2^r (2\lambda \lg p + r)$. This multiplication is then handled recursively.

Now, since $\lg p = O(\lg n)$, $\lg p \geqslant k$, $b \asymp \lg^2 n$ and $k = O(\lg n)$, we have $\lambda = (2 + O(1/\lg n)) b/\lg p$, and hence $n' = (4 + O(1/\lg \lg n)) b \, 2^r$, just as in section 6. The rest of the complexity analysis follows exactly as in the proof of Theorem 6.4, except for the computation of $p$ and $\omega$, which is considered below.

**Remark 8.1.** The role of the precision parameter $\lambda$ is to give some extra flexibility regarding the choice of $p$. If there was an efficient way to find a prime $p = 1 \pmod{2^k}$ larger than $2^{2b+k}$ (but not too much larger), and an efficient way to find a suitable $2^k$-th root of unity modulo $p$, then we could always take $\lambda := 1$ and obtain an algorithm working directly over the finite field $\mathbb{F}_p$.

## 8.2. Computing suitable $p$ and $\omega$

Given a transform length $2^k$ for $k \geqslant 1$, our aim is to find a prime $p$ such that $p = 1 \pmod{2^k}$, i.e., such that $2^k$ divides $p - 1$. Denote by $p_0(k)$ the smallest such prime.

Heath-Brown has conjectured that $p_0(k) = O(2^k k^2)$ [25], but given the current state of knowledge in number theory, we are only able to prove a result of the following type.

LEMMA 8.2. *For all sufficiently large $k$ we have $p_0(k) < 2^{6k}$, and we may compute $p_0(k)$ in time $O(2^{5k} k^{O(1)})$.*

**Proof.** This is a special case of Linnik's theorem [34, 35], which states that there exist constants $C$ and $L$ such that for any $a, b \in \mathbb{N}$ with $\gcd(a, b) = 1$, there exists a prime number $p = a \pmod{b}$ with $p < C b^L$. The best currently known estimate $L \leqslant 5.2$ for $L$ is due to Xylouris [53]. Applying this result for $a = 1$ and $b = 2^k$, we get the bound $p < 2^{6k}$ for large enough $k$. The complexity bound follows by testing $2^k + 1, 2 \cdot 2^k + 1, 3 \cdot 2^k + 1, \ldots$ for primality until we find $p$, using a polynomial time primality test [1]. □

The difficulty with this result — already noted in [15] — is that the time required to find $p$ greatly exceeds the time bound we are trying to prove for $\mathsf{I}(n)$!

To avoid this problem, De et al suggested using a multivariate splitting, i.e., by encoding each integer as a polynomial in $\mathbb{Z}[X_1, \ldots, X_m]$ for suitable $m$, say $m \geqslant 7$. One then uses $m$-dimensional DFTs to multiply the polynomials. Since the transform length is shorter, one can get away with a smaller $p$. Unfortunately, this introduces further zero-padding and leads to a larger value of $K$, ruining our attempt to achieve the bound $O(n \log n \, 8^{\log^* n})$.

On the other hand, we note that the problem only really occurs at the top recursion level. Indeed, at deeper recursion levels, there is *exponentially more time* available at the previous level to compute $p$. So one possible workaround is to use a different, sufficiently fast algorithm at the top level, such as Fürer's algorithm, and then switch to the algorithm sketched in section 8.1 for the remaining levels. In this way one still obtains the bound $O(n \log n \, 8^{\log^* n})$, and asymptotically almost all of the computation is done using the algorithm of section 8.1.

If one insists on avoiding $\mathbb{C}$ entirely, there are still many choices: one could use the algorithm of De et al at the top level, or use a multivariate version of the algorithm of section 8.1. One could even use the Schönhage–Strassen algorithm, whose main recursive step yields the bound $\mathsf{I}(n) = O(n^{1/2} \mathsf{I}(n^{1/2}) + n \log n)$; applying this three times gives $\mathsf{I}(n) = O(n^{7/8} \mathsf{I}(n^{1/8}) + n \log n)$, and then to multiply integers with $n^{1/8}$ bits, one can find a suitable prime using Lemma 8.2 in time $O(n^{3/4 + o(1)}) = O(n)$.

Another way to work around the problem is to assume the generalised Riemann hypothesis (GRH). De et al pointed out that under GRH, it is possible to find a suitable prime efficiently using a randomised algorithm. Here we show that, under GRH, we can even use deterministic algorithms.

LEMMA 8.3. *Assume GRH. Then $p_0(k) = O(2^{2k} k^2)$, and we may compute $p_0(k)$ in time $O(2^k k^{O(1)})$.*

**Proof.** The first bound is given in [26], and the complexity bound follows similarly to the proof of Lemma 8.2. □

To use this result, we must modify the algorithm of section 8.1 slightly. Choose a constant $C > 3$ so that we can compute $p_0(k)$ in time $O(2^k k^C)$, as in Lemma 8.3. Increase the coefficient size from $(\lg n)^2$ to $(\lg n)^{C-1}$, and change the definition of admissibility accordingly. The transform length then decreases to $2^k = O(n / (\lg n)^{C-1})$, and the cost of computing $p$ decreases to only $O(n \lg n)$. The rest of the complexity analysis is essentially unchanged; the result is an algorithm with complexity $O(n \log n \, 8^{\log^* n})$, working entirely with modular arithmetic, in which the top recursion level does not need any special treatment.

Finally, we consider the computation of a suitable approximation to a $2^k$-th root of unity in $\mathbb{Q}_p$.

LEMMA 8.4. *Given $k, \lambda \geqslant 1$ and a prime $p = 1 \pmod{2^k}$, we may find $\tilde{\omega} \in \mathbb{Z} / p^\lambda \mathbb{Z}$ such that $\tilde{\omega} = \omega \pmod{p^\lambda}$ for some primitive $2^k$-th root of unity $\omega \in \mathbb{Q}_p$, in time $O(p^{1/4+\epsilon} + (k\,\lambda \log p)^{1+\epsilon})$ for any $\epsilon > 0$.*

**Proof.** We may find a generator $g$ of $(\mathbb{Z} / p \mathbb{Z})^*$ deterministically in time $O(p^{1/4+\epsilon})$ [48]. Then $\tilde{\omega}_0 = g^{(p-1)/2^k}$ is a primitive $2^k$-th root of unity in $\mathbb{Z}/p\mathbb{Z}$, and there is a unique primitive $2^k$-th root of unity $\omega \in \mathbb{Q}_p$ congruent to $\tilde{\omega}_0$ modulo $p$. Given $\tilde{\omega}_0$, we may compute $\omega \pmod{p^\lambda}$ using fast Newton lifting in time $O((k\,\lambda \log p)^{1+\epsilon})$ [9, Section 12.3]. □

In the context of section 8.1, we may assume that $\lambda = O((\lg n)^{O(1)})$ and $k = O(\lg n)$, so the cost of finding $\omega$ is $O(p^{1/4+\epsilon})$. This is certainly less than the cost of finding $p$ itself, using either Lemma 8.2 or Lemma 8.3.

## 9. Conjecturally faster multiplication

It is natural to ask whether the approaches from sections 6, 7 or 8 can be further optimised, to obtain a complexity bound $\mathsf{I}(n) = O(n \log n\, K^{\log^* n})$ with $K < 8$.

In Fürer's algorithm, the complexity is dominated by the cost of multiplications in $R = \mathbb{C}[X] / (X^{2^{r-1}} + 1)$. If we could use a similar algorithm for a much simpler $R$, then we might achieve a better bound. Such an algorithm was actually given by Fürer [17], under the assumption that there exist sufficiently many Fermat primes, i.e., primes of the form $F_m = 2^{2^m} + 1$. More precisely, his algorithm requires that there exists a positive integer $k$ such that for every $m \in \mathbb{N}$, the sequence $F_{m+1}, \ldots, F_{2^{m+k}}$ contains a prime number. The DFTs are then computed directly over $R = \mathbb{F}_{F_m}$ for suitable $m$, taking advantage of the fact that $\mathbb{F}_{F_m}$ contains a fast $2^{m+1}$-th primitive root of unity (namely the element 2) as well as a $2^{2^m}$-th primitive root of unity. It can be shown that a suitably optimised version of this hypothetical algorithm achieves $K = 4$: we still pay a factor of two due to the fact that we compute both forward and inverse transforms, and we pay another factor of two for the zero-padding in the recursive reduction. Unfortunately, it is likely that $F_4 = 65537$ is the last Fermat prime [13].

In the $K = 8$ algorithm of section 6, a potential bottleneck arises during the short transforms, when we use Kronecker substitution to multiply polynomials in $\mathbb{C}_p[X] / (X^{2^r} - 1)$. We really only need the high $p$ bits of each coefficient of the product (i.e., of the real and imaginary parts), but we are forced to allocate roughly $2p$ bits per coefficient in the Kronecker substitution, and then we discard roughly half of the output. This problem is similar to the well-known obstruction that prevents us from using FFT methods to compute a "short product", i.e., the high $n$ bits or low $n$ bits of the product of two $n$-bit integers, any faster than computing the full $2n$ bits.

In this section, we present a variant of the algorithm of section 6, in which the coefficient ring $\mathbb{C}$ is replaced by a finite field $\mathbb{F}_p[i]$, where $p = 2^q - 1$ is a Mersenne prime. Thus "short products" are replaced by "cyclic products", namely by multiplications modulo $2^q - 1$. This saves a factor of two at each recursion level, and consequently reduces $K$ from 8 to 4.

This change of coefficient ring introduces several technical complications. First, it is of course unknown if there are infinitely many Mersenne primes. Thus we are forced to rely on unproved conjectures about the distribution of Mersenne primes.

Second, $q$ is always prime (except possibly at the top recursion level). Thus we cannot cut up an element of $\mathbb{Z} / p \mathbb{Z}$ into equal-sized chunks with an integral number of bits, and still expect to take advantage of cyclic products. In other words, $q$ is very far from being admissible in the sense of section 6. To work around this, we deploy a variant of an algorithm of Crandall and Fagin [12], which allows us to work with chunks of varying size. The Crandall–Fagin algorithm was originally presented over $\mathbb{C}$, and depended crucially on the fact that $\mathbb{R}$ contains suitable roots of 2. In our setting, we work over $\mathbb{F}_{p'}[i] \cong \mathbb{F}_{(p')^2}$, where $p' = 2^{q'} - 1$ is a Mersenne prime exponentially smaller than $p$. Happily, $\mathbb{F}_{p'}$ contains suitable roots of 2, and this enables us to adapt their algorithm to our setting. Moreover, since $(p')^2 - 1 = 2^{q'+1}(2^{q'-1} - 1)$, the field $\mathbb{F}_{p'}[i]$ contains roots of unity of high power-of-two order, namely of order $2^{q'+1}$, so we can perform FFTs over $\mathbb{F}_{p'}[i]$ very efficiently.

Finally, we can no longer use Kronecker substitution, as this would reintroduce the very zero-padding we are trying to avoid. Instead, we take our basic problem to be *polynomial* multiplication over $(\mathbb{Z}/p\mathbb{Z})[\mathrm{i}]$ (where $p = 2^q - 1$ is not necessarily prime). After the Crandall–Fagin splitting step, we have a *bivariate* multiplication problem over $\mathbb{F}_{p'}[\mathrm{i}]$, which is solved using 2-dimensional FFTs over $\mathbb{F}_{p'}[\mathrm{i}]$. These FFTs are in turn reduced to 1-dimensional FFTs using standard methods; this dimension reduction is, roughly speaking, the analogue of Kronecker substitution in this algorithm. (Indeed, it is also possible to give an algorithm along these lines that works over $\mathbb{C}$ but avoids Kronecker substitution entirely; this still yields $K = 8$ because of the "short product" problem mentioned above.) For the 1-dimensional transforms, we use the same technique as in previous sections: we use Cooley–Tukey's algorithm to decompose them into "short transforms" of exponentially shorter length, then use Bluestein's method to convert them to (univariate) polynomial products, and finally evaluate these products recursively.

### 9.1. Mersenne primes

Let $\pi_m(x)$ denote the number of Mersenne primes less than $x$. Based on probabilistic arguments and numerical evidence, Lenstra, Pomerance and Wagstaff have conjectured that

$$\pi_m(x) \;\sim\; \frac{\mathrm{e}^\gamma}{\log 2} \log \log x$$

as $x \to \infty$, where $\gamma = 0.5772...$ is the Euler constant [52, 40]. Our fast multiplication algorithm relies on the following slightly weaker conjecture.

CONJECTURE 9.1. *There exist constants $0 < a < b$ such that for all $x > 3$,*

$$a \log \log x < \pi_m(x) < b \log \log x.$$

PROPOSITION 9.2. *Assume Conjecture 9.1 and let $c := b/a$. For any integer $n \geqslant 2$, there exists a Mersenne prime $p = 2^q - 1$ in the interval $2^n < p < 2^{n^c}$. Given $n$, we may compute the smallest such $p$, and find a primitive $2^{q+1}$-th root of unity in $\mathbb{F}_p[\mathrm{i}]$, in time $O(n^{(3+o(1))c})$.*

**Proof.** The required prime exists since for $n \geqslant 2$ we have

$$\pi_m(2^{n^c}) > a \log \log (2^{n^c}) = a c \log n + a \log \log 2 > b \log n + b \log \log 2 = b \log \log(2^n) > \pi_m(2^n).$$

An integer of the form $2^q - 1$ may be tested for primality in time $q^{2+o(1)}$ using the Lucas–Lehmer primality test [13]. A simple way to compute $p$ is to apply this test successively for all $q \in \{n+1, ..., \lfloor n^c \rfloor\}$; this takes time $O(n^{(3+o(1))c})$. A primitive $2^{q+1}$-th root of unity $\omega$ may be computed by the formula $\omega := 2^{2^{q-2}} + (-3)^{2^{q-2}} \mathrm{i} \in \mathbb{F}_p[\mathrm{i}]$ in time $O(q^{2+o(1)})$; see [43] or [14, Corollary 5]. □

### 9.2. Crandall and Fagin's algorithm revisited

Let $p = 2^q - 1$ be a Mersenne number (not necessarily prime). The main integer multiplication algorithm depends on a variant of Crandall and Fagin's algorithm that reduces multiplication in $(\mathbb{Z}/p\mathbb{Z})[\mathrm{i}][X]/(X^M - 1)$ to multiplication in $\mathbb{F}_{p'}[\mathrm{i}][X,Y]/(X^M - 1, Y^N - 1)$, where $p' = 2^{q'} - 1$ is a suitably smaller Mersenne prime (assuming that such a prime exists).

To explain the idea of this reduction, we first consider the simpler univariate case, in which we reduce multiplication in $(\mathbb{Z}/p\mathbb{Z})[\mathrm{i}]$ to multiplication in $\mathbb{F}_{p'}[\mathrm{i}][Y]/(Y^N - 1)$. Here we require that $N \leqslant q$, that $\gcd(N, q') = 1$ and that $q' \geqslant 2 \lceil q/N \rceil + \lg N + 3$. For any $k \in \mathbb{N}$, we will write $\mathbb{N}_k = \{0, ..., k-1\}$ and $\mathbb{Z}_k = \{-(k-1), ..., k-1\}$.

Assume that we wish to compute the product of $u, v \in (\mathbb{Z}/p\mathbb{Z})[\mathrm{i}]$. Considering $u$ and $v$ as elements of $\mathbb{N}_p[\mathrm{i}]$ modulo $p$, we decompose them as

$$u = \sum_{i=0}^{N-1} u_i\, 2^{e_i}, \qquad v = \sum_{i=0}^{N-1} v_i\, 2^{e_i}, \qquad (9.1)$$

where

$$\begin{aligned} e_i &:= \lceil q\, i/N \rceil, \\ u_i, v_i &\in \mathbb{N}_{2^{e_{i+1}-e_i}}[\mathrm{i}]. \end{aligned}$$

We regard $u_i$ and $v_i$ as complex "digits" of $u$ and $v$, where the base $2^{e_{i+1}-e_i}$ varies with the position $i$. Notice that $e_{i+1} - e_i$ takes only two possible values: $\lfloor q/N \rfloor$ or $\lceil q/N \rceil$.

For $0 \leqslant i < N$, let

$$c_i := Ne_i - qi, \qquad (9.2)$$

so that $0 \leqslant c_i < N$. For any $0 \leqslant i_1, i_2 < N$, define $\delta_{i_1,i_2} \in \mathbb{Z}$ as follows. Choose $\sigma \in \{0, 1\}$ so that $i := i_1 + i_2 - \sigma N$ lies in the interval $0 \leqslant i < N$, and put

$$\delta_{i_1,i_2} := e_{i_1} + e_{i_2} - e_i - \sigma q.$$

From (9.2), we have

$$c_{i_1} + c_{i_2} - c_i = N(e_{i_1} + e_{i_2} - e_i) - q(i_1 + i_2 - i) = N\delta_{i_1,i_2}.$$

Since the left hand side lies in the interval $(-N, 2N)$, this shows that $\delta_{i_1,i_2} \in \{0, 1\}$. Now, since $2^q = 1 \pmod{p}$ and $e_{i_1} + e_{i_2} = e_i + \delta_{i_1,i_2} \pmod{q}$, we have

$$uv = \sum_{i_1=0}^{N-1} \sum_{i_2=0}^{N-1} u_{i_1} v_{i_2} 2^{e_{i_1}+e_{i_2}} = \sum_{i=0}^{N-1} w_i 2^{e_i} \pmod{p},$$

where

$$w_i := \sum_{i_1+i_2 = i \,(\mathrm{mod}\, N)} 2^{\delta_{i_1,i_2}} u_{i_1} v_{i_2}.$$

Since $|u_{i_1}| < \sqrt{2} \cdot 2^{\lceil q/N \rceil}$ and similarly for $v_{i_2}$, we have $w_i \in \mathbb{Z}_{4\lceil q/N \rceil + 1 N}[i]$. Note that we may recover $uv$ from $w_0, \ldots, w_{N-1}$ in time $O(q)$, by a standard overlap-add procedure (provided that $N = O(q/\lg q)$).

Let $h$ be the inverse of $q'$ modulo $N$; this inverse exists since we assumed $\gcd(N, q') = 1$. Let $\theta := 2^h \in \mathbb{F}_{p'}$, so that

$$\theta^N = 2^{hN} = 2,$$

since 2 has order $q'$ in $\mathbb{F}_{p'}$. The quantity $\theta$ plays the same role as the real $N$-th root of 2 appearing in Crandall–Fagin's algorithm.

Now define polynomials $U, V \in \mathbb{F}_{p'}[i][Y]/(Y^N - 1)$ by $U_i := \theta^{c_i} u_i$ and $V_i := \theta^{c_i} v_i$ for $0 \leqslant i < N$, and let $W = W_0 + \cdots + W_{N-1} Y^{N-1} := UV$ be their (cyclic) product. Then

$$\tilde{w}_i := \theta^{-c_i} W_i = \sum_{i_1+i_2 = i \,(\mathrm{mod}\, N)} \theta^{-c_i} U_{i_1} V_{i_2} = \sum \theta^{c_{i_1}+c_{i_2}-c_i} u_{i_1} v_{i_2} = \sum 2^{\delta_{i_1,i_2}} u_{i_1} v_{i_2}$$

coincides with the reinterpretation of $w_i$ as an element of $\mathbb{F}_{p'}[i]$. Moreover, we may recover $w_i$ unambiguously from $\tilde{w}_i$, as $q' \geqslant 2\lceil q/N \rceil + \lg N + 3$ and $w_i \in \mathbb{Z}_{4\lceil q/N \rceil + 1 N}[i]$. Altogether, this shows how to reduce multiplication in $(\mathbb{Z}/p\mathbb{Z})[i]$ to multiplication in $\mathbb{F}_{p'}[i][Y]/(Y^N - 1)$.

**Remark 9.3.** The pair $(e_{i+1}, c_{i+1})$ can be computed from $(e_i, c_i)$ in $O(\lg q)$ bit operations, so we may compute the sequences $e_0, \ldots, e_{N-1}$ and $c_0, \ldots, c_{N-1}$ in time $O(N \lg q)$. Moreover, since $c_{i+1} - c_i$ takes on only two possible values, we may compute the sequence $\theta^{c_0}, \ldots, \theta^{c_{N-1}}$ using $O(N)$ multiplications in $\mathbb{F}_{p'}[i]$.

## 9.3. Bivariate Crandall–Fagin reduction

Generalising the discussion of the previous section, we now show how to reduce multiplication in $(\mathbb{Z}/p\mathbb{Z})[i][X]/(X^M - 1)$, for a given $M \geqslant 1$, to multiplication in $\mathbb{F}_{p'}[i][X, Y]/(X^M - 1, Y^N - 1)$. For this, we require that $N \leqslant q$, that $\gcd(N, q') = 1$ and that $q' \geqslant 2\lceil q/N \rceil + \lg(MN) + 3$.

Indeed, consider two cyclic polynomials $u = u_0 + \cdots + u_{M-1} X^{M-1}$ and $v = v_0 + \cdots + v_{M-1} X^{M-1}$ in $(\mathbb{Z}/p\mathbb{Z})[i][X]/(X^M - 1)$. We cut each of the coefficients $u_i, v_i \in (\mathbb{Z}/p\mathbb{Z})[i]$ into $N$ chunks $u_{i,j}$ and $v_{i,j}$ of bit size at most $\lceil q/N \rceil$, using the same varying base strategy as above. With $\theta^N = 2$ and $c_j$ as before, we next form the bivariate cyclic polynomials

$$U := \sum_{i,j} u_{i,j} \theta^{c_j} X^i Y^j, \qquad V := \sum_{i,j} v_{i,j} \theta^{c_j} X^i Y^j$$

in $\mathbb{F}_{p'}[\mathrm{i}][X,Y]/(X^M-1, Y^N-1)$. Setting

$$W := UV = \sum_{i,j} w_{i,j} \theta^{c_j} X^i Y^j,$$

the same arguments as in the previous section yield

$$w_{i,j} = \sum_{i_1+i_2=i \,(\mathrm{mod}\, M)} \sum_{j_1+j_2=j \,(\mathrm{mod}\, N)} 2^{\delta_{j_1,j_2}} u_{i_1,j_1} v_{i_2,j_2}.$$

Using the assumption that $q' \geqslant 2 \lceil q/N \rceil + \lg(MN) + 3$, we recover the coefficients $w_{i,j}$, and hence the product $uv$, from the bivariate cyclic convolution product $W = UV$.

### 9.4. Conjecturally faster multiplication

Let $q \geqslant 2$ and $p := 2^q - 1$ (not necessarily prime). We will take our basic recursive problem to be multiplication in $(\mathbb{Z}/p\mathbb{Z})[\mathrm{i}][X]/(X^M-1)$ for suitable $M$. We need $M$ somewhat larger than $q$; this is analogous to the situation in section 6, where we chose a "short transform length" somewhat larger than the coefficient size. Thus we set $M = M(q) := 2^{\mu(q)}$ where $\mu(q)$ is defined as follows.

LEMMA 9.4. *There exists an increasing function $\mu : \mathbb{N} \to \mathbb{N}$ such that*

$$0 \leqslant \mu(q) - (\log_2 q)(\log_2 \log_2 q) \leqslant 2 \tag{9.3}$$

*for all $q \geqslant 2$, and such that we may compute $\mu(q)$ in time $(\log q)^{1+o(1)}$.*

**Proof.** Let $f(q) := (\log_2 q)(\log_2 \log_2 q)$. Using [6], we may construct a function $g(q)$ such that $|g(q) - f(q)| \leqslant 1/q$ for all $q \geqslant 2$, and which may be computed in time $(\log q)^{1+o(1)}$. One checks that $f(q+1) - f(q) \geqslant 2/q$ for all $q \geqslant 2$, so $g(q+1) \geqslant f(q+1) - \frac{1}{q+1} \geqslant f(q+1) - \frac{1}{q} \geqslant f(q) + \frac{1}{q} \geqslant g(q)$ for $q \geqslant 2$. Thus $g(q)$ is increasing, and $\mu(q) := \lfloor g(q) + 3/2 \rfloor$ has the desired properties. $\square$

We say that an integer $n \geqslant 2$ is *admissible* if it is of the form $n = qM$ where $M := M(q)$ for some $q \geqslant 2$. (This should not be confused with the notion of admissibility of section 6.) An element of $(\mathbb{Z}/p\mathbb{Z})[\mathrm{i}][X]/(X^M-1)$ is then represented by $2n$ bits. Note that $q \mapsto qM(q)$ is strictly increasing, so there is a one-to-one correspondence between integers $q \geqslant 2$ and admissible $n$. For $x \geqslant 2$ we define $\beta(x)$ to be the smallest admissible integer $n \geqslant x$.

LEMMA 9.5. *We have $\beta(n) = O(n)$ as $n \to \infty$. Given $n \geqslant 2$, we may compute $\beta(n)$, and the corresponding $q$, in time $o(n)$.*

**Proof.** From (9.3) we have $(q+1)M(q+1)/(qM(q)) = O(2^{\mu(q+1)-\mu(q)}) = O(1)$; this immediately implies that $\beta(n) = O(n)$.

Suppose that we wish to compute $\beta(n)$ for some $n$. We assume that $n$ is large enough that the definition $q_0 := 2^{\lceil \lg n/(\lg \lg n - \lg \lg \lg n - 1) \rceil}$ makes sense and so that $q_0 \geqslant 2$. One checks that $(\log_2 q_0)(\log_2 \log_2 q_0) \geqslant \lg n$, so $\mu(q_0) \geqslant \lg n$ and hence $q_0 M(q_0) \geqslant n$. To find the smallest suitable $q$, we may simply compute $qM(q)$ for each $q = 2, 3, \ldots, q_0$, and compare with $n$. This takes time $O(q_0 (\log q_0)^{1+o(1)}) = o(n)$. $\square$

Now let $q \geqslant 2$, $p := 2^q - 1$ and $M := M(q)$. Consider the problem of computing $t \geqslant 1$ products $u_1 v, \ldots, u_t v$ with $u_1, \ldots, u_t, v \in (\mathbb{Z}/p\mathbb{Z})[\mathrm{i}][X]/(X^M-1)$. We denote by $\mathsf{C}_t(n)$ the complexity of this problem, where $n := qM(q)$ is the admissible integer corresponding to $q$. As in section 6, we define $\mathsf{C}(n) := \sup_{t \geqslant 1} \mathsf{C}_t(n)/(2t+1)$.

Notice that multiplication of two integers of bit size $\leqslant k$ reduces to the above problem, for $t = 1$, via a suitable Kronecker segmentation. Indeed, let $n := \beta(8k) = qM(q)$ for some $q$, and encode the integers as integer polynomials of degree less than $M/2$ with coefficients of bit size $m := \lceil k/(M/2) \rceil$. The desired product may be recovered from the product in $(\mathbb{Z}/p\mathbb{Z})[\mathrm{i}][X]/(X^M-1)$, as

$$2m + \lg(M/2) \leqslant \frac{4k}{M} + \mu(q) \leqslant \frac{q}{2} + \mu(q) \leqslant q - 1$$

for large $q$. Thus, as in section 6, we have $\mathsf{I}(k) \leqslant 3\,\mathsf{C}(O(k)) + O(k)$, and it suffices to obtain a good bound for $\mathsf{C}(n)$.

Now suppose additionally that $p = 2^q - 1$ is *prime*. In this case $(\mathbb{Z}/p\mathbb{Z})[i] = \mathbb{F}_p[i]$ is a field, and as noted above, it contains $2^{q+1}$-th roots of unity, so we may define DFTs of length $2^r$ over $\mathbb{F}_p[i]$ for any $r \leqslant q + 1$. In particular, for $r \leqslant q$ we may use Bluestein's algorithm to compute DFTs of length $2^r$. Denote by $\mathsf{B}_{q,t}(2^r)$ the cost of evaluating $t$ independent DFTs of length $2^r$ over $\mathbb{F}_p[i]$, and put $\mathsf{B}_q(2^r) := \sup_{t \geqslant 1} \mathsf{B}_{q,t}(2^r)/(2t+1)$. Here we assume as usual that a $2^{r+1}$-th root of unity is known, and that the corresponding Bluestein root table has been precomputed.

Let us apply these definitions in the case $r := \lg M$; this is permissible, as $\lg M \leqslant q$ for sufficiently large $q$. Since convolution of length $M$ over $\mathbb{F}_p[i]$ is exactly the basic recursive problem, and since one of the operands is fixed, we have $\mathsf{B}_{q,t}(M) \leqslant \mathsf{C}_t(n) + O(tM\,\mathsf{I}(q))$, where $n := qM$, and hence

$$\mathsf{B}_q(M) \;\leqslant\; \mathsf{C}(n) + O(M\,\mathsf{I}(q)). \tag{9.4}$$

THEOREM 9.6. *Assume Conjecture 9.1. Then there exists $x_0 \geqslant 2$ and a logarithmically slow function $\Phi: (x_0, \infty) \to \mathbb{R}$ with the following property. For all admissible $n > x_0$, there exists an admissible $n' \leqslant \Phi(n)$ such that*

$$\frac{\mathsf{C}(n)}{n \lg n} \;\leqslant\; \left(4 + O\!\left(\frac{1}{\lg\lg\lg n}\right)\right) \frac{\mathsf{C}(n')}{n' \lg n'} + O(1). \tag{9.5}$$

**Proof.** Let $n := qM$ with $M = M(q)$. Assume that we wish to compute $t \geqslant 1$ products with one fixed operand. Our goal is to reduce to a problem of the same form, but for exponentially smaller $n$.

**Choose parameters.** Let $p' = 2^{q'} - 1$ be the smallest Mersenne prime larger than $2^{(\lg M)^2}$. By Proposition 9.2, we have $2^{(\lg M)^2} < p' < 2^{(\lg M)^{2c}}$, whence $(\lg M)^2 \leqslant q' \leqslant (\lg M)^{2c}$, for some absolute constant $c > 1$. Moreover, we may compute $p'$, together with a primitive $2^{q'+1}$-th root of unity $\omega$ in $\mathbb{F}_{p'}[i]$, in time $O((\lg M)^{(6+o(1))c}) = O(n \lg n)$. We define $M' := M(q')$ and $n' := q'M'$.

The algorithm must perform various multiplications in $\mathbb{F}_{p'}[i]$, at cost $O(\mathsf{I}(q'))$. For simplicity we will use Schönhage–Strassen's algorithm for these multiplications, i.e., we will take $\mathsf{I}(q') = O(q' \lg q' \lg \lg q')$. Since $\lg q' = O(\lg \lg M) = O(\lg \lg n)$, we have

$$\mathsf{I}(q') \;=\; O(q' \lg \lg n \lg \lg \lg n).$$

**Crandall–Fagin reduction.** We use the framework of section 9.3 to reduce the basic multiplication problem in $(\mathbb{Z}/p\mathbb{Z})[i][X]/(X^M - 1)$ to multiplication in $\mathbb{F}_{p'}[i][X,Y]/(X^M - 1, Y^N - 1)$ for suitable $N$. We take $N := 2^\ell s$ where

$$\begin{aligned}
\ell &:= \lg\!\left(\frac{2q}{q' \lg \lg q}\right), \\
s &:= 2\left\lceil \frac{q}{2^\ell (q' - \lg^2 q)} \right\rceil + 1.
\end{aligned}$$

We also write $L := 2^\ell$. The definition of $s$ makes sense for large $q$ since $q' \geqslant (\lg M)^2 \asymp (\lg q \lg \lg q)^2$. Let us check that the hypotheses of section 9.3 are satisfied for large $q$. We have $L \asymp q/(q' \lg \lg q)$ and hence $s \asymp \lg \lg q$; in particular, $s \neq q'$, so $\gcd(N, q') = 1$, and also $N \asymp q/q' \prec q/\lg q$. Since $N = Ls \geqslant 2q/(q' - \lg^2 q)$, we also have $2\lceil q/N \rceil \leqslant q' - \lg^2 q + O(1)$, and thus $2\lceil q/N \rceil + \lg(MN) + 3 \leqslant q'$ since $\lg(MN) = O(\lg q \lg \lg q)$.

We also note for later use the estimate

$$MNq' \;=\; \left(2 + O\!\left(\frac{1}{\lg \lg n}\right)\right) n.$$

Indeed, since $s \asymp \lg \lg q$ we have

$$s \;=\; \left(2 + O\!\left(\frac{1}{\lg \lg q}\right)\right) \frac{q}{2^\ell (q' - \lg^2 q)},$$

and we already noticed earlier that $(\lg^2 q)/q' = O(1/(\lg \lg q)^2) = O(1/\lg \lg q)$.

To assess the cost of the Crandall–Fagin reduction, we note that computing the $e_i$ and $c_i$ costs $O(N \lg q) = O(n \lg n)$ (see Remark 9.3), the splitting itself and final overlap-add phase require time $O(tn)$, and the various multiplications by $\theta$, $\theta^{c_i}$ and $\theta^{-c_i}$ have cost $O(tMN\,\mathsf{I}(q')) = O(tn\,\mathsf{I}(q')/q') = O(tn \lg n)$.

**Reduction to power-of-two lengths.** Next we reduce multiplication in $\mathbb{F}_{p'}[i][X,Y]/(X^M-1, Y^N-1)$ to multiplication in $\mathcal{R}[X,Z]/(X^M-1, Z^L-1)$, where $\mathcal{R} := \mathbb{F}_{p'}[i][U]/(U^s-1)$. In fact, since $\gcd(L,s) = 1$, these rings are isomorphic, via the map that sends $X$ to $X$ and $Y$ to $ZU$. Evaluating this isomorphism corresponds to rearranging the coefficients according to the rule $i \mapsto (i_0, i_1)$, where $i \in \{0, ..., N-1\}$ is the exponent of $Y$ and where $i_0 := i \bmod L$ and $i_1 := i \bmod s$ are the exponents of $Z$ and $U$. This may be achieved in time $O(tMN \lg N\,(q' + \lg N)) = O(t n \lg n)$ using the same sorting strategy as in section 2.3. The inverse rearrangement is handled similarly.

**Reduction to univariate transforms.** For multiplication in $\mathcal{R}[X,Z]/(X^M-1, Z^L-1)$, we will use bivariate DFTs over $\mathcal{R}$. This is possible because $\mathbb{F}_{p'}[i]$ contains both $M$-th and $L$-th primitive roots of unity, namely $\omega^{2^{q'+1}/M}$ and $\omega^{2^{q'+1}/L}$, since $q' \succ \lg M$ and $q' \succ \lg L$. More precisely, we must perform $t+1$ forward bivariate DFTs and $t$ inverse bivariate DFTs of length $M \times L$ over $\mathcal{R}$, and $tML$ multiplications in $\mathcal{R}$. Each bivariate DFT reduces further to $sM$ univariate DFTs of length $L$ over $\mathbb{F}_{p'}[i]$ (with respect to $Z$) and $sL$ univariate DFTs of length $M$ over $\mathbb{F}_{p'}[i]$ (with respect to $X$). Interspersed between these steps are various matrix transpose operations of total cost $O(t\,s\,ML \lg(sML)\,q') = O(t\,n\lg n)$, to enable efficient access to the "rows" and "columns" (see section 2.1).

Multiplications in $\mathcal{R}$ are handled by zero-padding, i.e., we first use Cooley–Tukey to multiply in $\mathbb{F}_{p'}[i][U]/(U^{2^{\lceil \lg s \rceil + 1}} - 1)$, and then reduce modulo $U^s - 1$. The total cost of these multiplications is $O(t\,ML\,s\,\lg s\,\mathsf{I}(q')) = O(t\,n\,\lg s\,\mathsf{I}(q')/q') = O(t\,n\,\lg\lg n\,(\lg\lg\lg n)^2) = O(t\,n\,\lg n)$.

**Reduction to short transforms.** Consider one of the "long" univariate DFTs of length $2^k \in \{M, L\}$ over $\mathbb{F}_{p'}[i]$. We decompose the DFT into "short" DFTs of length $M'$ as follows. Let $r := \lg M' = O(\lg\lg n \lg\lg\lg n)$ and $d := \lceil k/r \rceil = O(\lg n/(\lg\lg n \lg\lg\lg n))$, and write $k = r_1 + \cdots + r_d$ where $r_i := r$ for $1 \leq i \leq d-1$ and $r_d := k - (d-1)r \leq r$. We use the algorithm $\mathcal{A} := \mathcal{A}_1 \odot \cdots \odot \mathcal{A}_d$, where for $1 \leq i \leq d-1$ we take $\mathcal{A}_i$ to be the algorithm based on Bluestein's method (discussed immediately before (9.4)), and where $\mathcal{A}_d$ is the usual Cooley–Tukey algorithm over $\mathbb{F}_{p'}[i]$. Let $\mathsf{D}_k$ be the cost of a single invocation of $\mathcal{A}$ (or of the corresponding inverse transform $\mathcal{A}'$). By (2.4) we have

$$\mathsf{D}_k \;\leq\; (d-1)\,\mathsf{B}_{q',2^{k-r}}(2^r) + O(2^{k-r_d} 2^{r_d} r_d \mathsf{I}(q')) + O(d\,2^k \mathsf{I}(q')) + O(2^k q' \lg n).$$

The cost of precomputing the necessary root tables is only $O(2^k \mathsf{I}(q'))$. By definition $\mathsf{B}_{q',2^{k-r}}(2^r) \leq (2 \cdot 2^{k-r} + 1)\,\mathsf{B}_{q'}(M')$. From (9.4) and the estimate $2^{k-r} \succ \lg\lg n$, the first term becomes

$$(d-1)\,\mathsf{B}_{q',2^{k-r}}(2^r) \;\leq\; (2 + O(1/\lg\lg n))\,(d-1)\,2^{k-r}\,\mathsf{C}(n') + O(d\,2^{k-r} M' \mathsf{I}(q')).$$

The contribution to $\mathsf{D}_k$ from all terms involving $\mathsf{I}(q')$ is

$$O(2^k (r_d + d)\,\mathsf{I}(q')) = O\!\left(2^k \frac{\lg n}{\lg\lg n \lg\lg\lg n}\,q' \lg\lg n \lg\lg\lg n\right) = O(2^k q' \lg n),$$

so

$$\mathsf{D}_k \;\leq\; (2 + O(1/\lg\lg n))\,(d-1)\,2^{k-r}\,\mathsf{C}(n') + O(2^k q' \lg n).$$

Denoting by $\mathsf{D}$ the cost of a bivariate DFT of length $M \times L$ over $\mathcal{R}$, we thus have (ignoring the transposition costs, which were included earlier)

$$\begin{aligned}
\mathsf{D} \;&=\; sL\,\mathsf{D}_{\lg M} + sM\,\mathsf{D}_{\lg L} \\
&\leq\; \left(2 + O\!\left(\frac{1}{\lg\lg n}\right)\right)\!\left(sL\left\lfloor\frac{\lg M}{\lg M'}\right\rfloor\frac{M}{M'} + sM\left\lfloor\frac{\lg L}{\lg M'}\right\rfloor\frac{L}{M'}\right)\mathsf{C}(n') + O(sLMq'\lg n) \\
&\leq\; \left(2 + O\!\left(\frac{1}{\lg\lg n}\right)\right)sLM\,\frac{\lg(LM)}{M' \lg M'}\,\mathsf{C}(n') + O(sLMq'\lg n) \\
&\leq\; \left(4 + O\!\left(\frac{1}{\lg\lg n}\right)\right)\frac{n \lg n}{n' \lg M'}\,\mathsf{C}(n') + O(n \lg n).
\end{aligned}$$

Moreover, since

$$\frac{\lg n'}{\lg M'} = 1 + \frac{\lg q'}{\lg M'} = 1 + O\!\left(\frac{1}{\lg\lg q'}\right) = 1 + O\!\left(\frac{1}{\lg\lg\lg n}\right),$$

we get

$$\mathsf{D} \;\leq\; \left(4 + O\!\left(\frac{1}{\lg\lg\lg n}\right)\right)\frac{n \lg n}{n' \lg n'}\,\mathsf{C}(n') + O(n \lg n).$$

We must perform $2t+1$ bivariate DFTs; the bound (9.5) then follows exactly as in the proof of Theorem 6.4.

For large $n$, we have $\log q' = O(\log \log M) = O(\log \log n)$, so $\log n' = \log q' + O(\mu(q')) = O(\log q' \log \log q') = O(\log \log n \log \log \log n)$. Thus there exists a constant $C > 0$ such that $\log \log \log n' \leqslant \log \log \log \log n + C$ for large $n$, and we may take $\Phi(x) := \exp^{\circ 3}(\log^{\circ 4} x + C)$. □

**Proof of Theorem 1.2.** Follows from Theorem 9.6 and Proposition 5.3, analogously to the proof of Theorem 1.1. □